\documentclass[11pt]{article}
\usepackage{jcappub}
\usepackage{bm}
\usepackage{color}
\usepackage{graphicx}

\newcommand{\be}{\begin{equation}}
\newcommand{\ee}{\end{equation}}
\newcommand{\een}{\end{subequations}}
\newcommand{\ben}{\begin{subequations}}
\newcommand{\beq}{\begin{eqalignno}}
\newcommand{\eeq}{\end{eqalignno}}

\newcommand{\lsim}{\mathrel{\mathop{\kern 0pt \rlap
      {\raise.2ex\hbox{$<$}}}\lower.9ex\hbox{\kern-.190em $ \sim$}}}
\newcommand{\gsim}{\mathrel{\mathop{\kern 0pt
      \rlap{\raise.2ex\hbox{$>$}}}\lower.9ex\hbox{\kern-.190em $\sim$}}}

\title{From direct detection to relic abundance: the case of
  proton--philic spin--dependent inelastic Dark Matter}

\author{Stefano Scopel\footnote{Temporary address: Instituto de F\'isica Te\'orica (UAM/CSIC),
  Universidad Aut\'onoma de Madrid, Cantoblanco, 28049, Madrid,
  Spain},}
\author{Hyeonhye Yu}
\emailAdd{scopel@sogang.ac.kr}
\emailAdd{skyh2yu@gmail.com}

\affiliation{Department of Physics, Sogang University, Seoul, South
  Korea}

\abstract{We discuss strategies to make inferences on the thermal
  relic abundance of a Weakly Interacting Massive Particle (WIMP) when
  the same effective dimension--six operator that explains an
  experimental excess in direct detection is assumed to drive
  decoupling at freeze--out, and apply them to the explicit scenario
  of WIMP inelastic up--scattering with spin--dependent couplings to
  protons (proton--philic Spin--dependent Inelastic Dark Matter,
  pSIDM), a phenomenological set--up containing two Dark Matter (DM)
  particles $\chi_1$ and $\chi_2$ with masses $m_{\chi}$=$m_{\chi_1}$
  and $m_{\chi_2}$=$m_{\chi}+\delta$ that we have shown in a previous
  paper to explain the DAMA effect in compliance with the constraints
  from other detectors. We also update experimental constraints on
  pSIDM, extend the analysis to the most general spin--dependent
  momentum--dependent interactions allowed by non--relativistic
  Effective Field Theory (EFT), and consider for the WIMP velocity
  distribution in our Galaxy $f(v)$ both a halo--independent approach
  and a standard Maxwellian. Under these conditions we find that the
  DAMA effect can be explained in terms of the particle $\chi_1$ in
  compliance with all the other constraints for all the analyzed EFT
  couplings and also for a Maxwellian $f(v)$. As far as the relic
  abundance is concerned, we show that the problem of calculating it
  by using direct detection data to fix the model parameters is
  affected by a strong sensitivity on $f(v)$ and by the degeneracy
  between the WIMP local density $\rho_{\chi}$ and the WIMP--nucleon
  scattering cross section, since $\rho_{\chi}$ must be rescaled with
  respect to the observed DM density in the neighborhood of the Sun
  when the calculated relic density $\Omega$ is smaller than the
  observed one $\Omega_0$. As a consequence, a DM direct detection
  experiment is not directly sensitive to the physical cut--off scale
  of the EFT, but on some dimensional combination that does not depend
  on the actual value of $\Omega$. However, such degeneracy can be
  used to develop a consistency test on the possibility that the WIMP
  is a thermal relic in the first place. When we apply it to the pSIDM
  scenario we find that only a WIMP with the standard spin--dependent
  interaction ${\cal O}$=$\bar{\chi}_1\gamma^{\mu}\gamma^5\chi_2
  \bar{q}\gamma_{\mu}\gamma^5 q$ +h.c. with quarks can be a thermal
  relic for, approximately, 10 GeV $\lsim m_{\chi}\lsim$ 16 GeV, 17
  keV$\lsim \delta\lsim$28 keV, and a large uncertainty on $\Omega$,
  $6\times 10^{-7}\Omega_0\lsim \Omega \lsim \Omega_0$. In order for
  the scenario to work the WIMP galactic velocity distribution must
  depart from a Maxwellian. Moreover, all the $\chi_2$ states must
  have already decayed today, and this requires some additional
  mechanism besides that provided by the ${\cal O}$ operator.}

\begin{document}

\maketitle

\section{Introduction}
\label{sec:introduction}

Weakly Interacting Massive Particles (WIMPs) are considered the most
natural candidates to provide the Dark Matter (DM) in the halo of our
Galaxy, and the search for their recoils off nuclear targets
represents the most direct way to detect them. In lack of any
experimental evidence of physics beyond the Standard Model at the
Large Hadron Collider (LHC) the use of non--relativistic Effective
Field Theories (EFT)\cite{haxton1,haxton2} to analyze WIMP direct
search data has become popular in the
literature\cite{buckley,catena_gondolo,catena_ibarra}.  In presence of
a putative signal, the question arises of how to shed light on the
ultraviolet completion of the EFT, or, at least, of how to use a given
EFT to find correlations among different types of WIMP
signals. Indeed, this has been discussed in the context of mono--jet
signal at the LHC\cite{eft_monojet1,eft_monojet2}, showing that the
ensuing constraints at low WIMP masses can be competitive with those
from direct searches. Such inferences however imply an extrapolation
from the scale of WIMP-nucleus elastic scattering (keV) to the LHC
scale (TeV), that becomes questionable if the relevant value of the
EFT cut--off scale does not significantly exceed the momentum transfer
of the involved physical
processes\cite{lhc_eft_validity1,lhc_eft_validity2,lhc_eft_validity3}.
Similarly, once an effective interaction operator is singled out from
direct detection data the question arises if the same operator can
yield an acceptable relic abundance through thermal
decoupling\cite{buckley}. Due to the non--relativistic nature of the
latter process in this case the involved energy extrapolation is much
smaller, since the energy scale relevant to WIMP annihilation at
freeze--out is of order $2 m_{\chi}$, with $m_{\chi}={\cal O}(GeV)$
the WIMP mass. As a consequence, the use of EFT to calculate the relic
abundance appears more reliable, and represents the first necessary
step upward in the bottom--up quest to shed light on the ultraviolet
completion of the EFT.

In the present paper we wish to address this latter issue, starting
from an explicit class of EFT low--energy models that we introduced in
a previous paper\cite{noi_idm_spin} and suggested by present
experimental data . Specifically, our goal is to develop operative
bottom--up strategies to test the possibility that the same effective
dimension--six operator that explains an experimental excess in WIMP
direct detection (at the keV scale) can also yield an acceptable relic
abundance through thermal decoupling (at the GeV scale). In
particular, we will also address this issue when the WIMP velocity
distribution in our Galaxy is not fixed.

Many underground experiments using different materials are strongly
pursuing the task of WIMP direct detection, with an impressive
improvement of the experimental sensitivities but somewhat confusing
and apparently contradictory results. In particular, the DAMA
experiment\cite{dama} has been measuring for more than 15 years a
yearly modulation effect with a sodium iodide target, consistent with
that expected from the elastic scattering of WIMPs due to the Earth
rotation around the Sun. However, many experimental collaborations
using nuclear targets different from $NaI$ and various
background--subtraction techniques to look for WIMP--elastic
scattering
(LUX\cite{lux,lux_2015_reanalysis,lux_complete},PANDA\cite{panda_run_8,panda_run_9}
XENON100\cite{xenon100}, XENON10\cite{xenon10},
KIMS\cite{kims,kims_modulation,kims2}, CDMS-$Ge$\cite{cdms_ge},
CDMSlite \cite{cdms_lite}, SuperCDMS\cite{super_cdms}, CDMS
II\cite{cdms_2015}, SIMPLE\cite{simple}, COUPP\cite{coupp},
PICASSO\cite{picasso,picasso_final}, PICO-2L\cite{pico2l,pico2l_2016},
PICO-60\cite{pico60}) have failed to observe any anomaly so far. This
implies severe constraints on the most popular WIMP scenarios used to
explain the DAMA excess: a WIMP with scattering cross section off
nuclei proportional to the square of the atomic mass number of the
target or to the target nuclear spin, and with a galactic velocity
distribution given by a Maxwellian. However, when the latter
assumptions are relaxed it is possible to show that compatibility
between an interpretation of the DAMA effect in terms of a WIMP signal
and constraints from other experiment can be recovered
\cite{noi_roma_2015,noi1,noi2,noi_eft_spin, noi_idm_spin}.

In particular, in the proton--philic Spin--dependent Inelastic Dark
Matter (pSIDM) scenario introduced in \cite{noi_idm_spin} the DAMA
modulation effect is explained by a WIMP which upscatters
inelastically\cite{inelastic} to a heavier state and predominantly
couples to the spin of protons. In such scenario constraints from
xenon and germanium targets are evaded dynamically, due to the
suppression of the WIMP coupling to neutrons, while those from
fluorine targets are evaded kinematically, because the minimal WIMP
incoming speed required to trigger upscatters off fluorine exceeds the
maximal WIMP velocity in the Galaxy, or is very close to it.  The
scenario of Ref.\cite{noi_idm_spin} is purely phenomenological, and
can be mainly considered as a proof of concept of the fact that the
parameter space of WIMP direct detection is wider that the one usually
assumed. However, it has also some attractive properties: for
instance, it predicts in a natural way large modulation fractions
compatible to the DAMA experimental data, and provides an explanation
of the maximum in the energy spectrum of the modulation amplitude
detected by DAMA in terms of WIMPs whose minimal incoming speed
matches the kinematic threshold for inelastic upscatters.

In light of this, in the present paper we wish to elaborate more on
such scenario, extending the discussion of Ref.\cite{noi_idm_spin} in
several directions: i) we update experimental constraints including
the latest results from LUX\cite{lux_2015_reanalysis,lux_complete},
PANDA\cite{panda_run_8,panda_run_9} and PICO-2L\cite{pico2l_2016}; ii)
we extend the discussion to the most general class of spin--dependent
interactions \cite{noi_eft_spin} making use of non--relativistic
EFT\cite{haxton1,haxton2}; iii) we discuss the thermal relic density
of such particle when its interaction with ordinary matter at the
scale of WIMP annihilation in the early Universe is described by the
same dimension--six operator responsible for elastic scattering.

The paper is organized as follows: in Section
\ref{sec:spin_idm_scenario} we summarize the main features of the
pSIDM scenario, extending it to the most general cases of a
spin--dependent coupling (as listed in Table \ref{table:ref_nref}); in
Section \ref{sec:relic} we discuss the DM relic density calculation;
in Section \ref{sec:bottom_up} we discuss the kind of inferences that
an excess in direct detection data can allow on the EFT cut--off
scale, showing that the effect of the rescaling of the local density
when the DM particle turns out to be subdominant in the Universe
limits our access on the latter, but allows to develop a robust
halo--independent consistency check on the possibility that the the
same effective DM model that explains the excess can fix the thermal
density; in Section \ref{sec:analysis} we show the results of our
numerical analysis for pSIDM in the specific class of effective models
introduced in Section \ref{sec:spin_idm_scenario}, and in Section
\ref{sec:conclusions} we provide our conclusions. In Appendix
\ref{app:sigmav} we provide the analytic expressions for the thermal
average of the WIMP annihilation cross section entering the relic
abundance calculation; in Appendix \ref{app:matrix_elements} we
summarize the procedure to relate the WIMP--nucleon coupling constants
entering direct detection to the WIMP--quark couplings that appear in
the calculation of the annihilation cross section; finally, in
Appendix \ref{app:exp} we provide the details of our treatment of the
latest experimental constraints from
LUX\cite{lux_2015_reanalysis,lux_complete},
PANDA\cite{panda_run_8,panda_run_9} and PICO-2L\cite{pico2l_2016}
(constraints from earlier experiments are implemented as described in
Refs. \cite{noi_eft_spin,noi_idm_spin}).

\section{The pSIDM scenario and DAMA}
\label{sec:spin_idm_scenario}

In this Section we briefly summarize the features of the scenario
introduced in Ref.\cite{noi_idm_spin} (we refer the reader to such
paper for further details).

The most stringent bounds on an interpretation of the DAMA effect in
terms of WIMP--nuclei scatterings are obtained by detectors using
xenon (LUX\cite{lux,lux_2015_reanalysis,lux_complete},
PANDA\cite{panda_run_8,panda_run_9}) and germanium
(CDMS\cite{cdms_ge,cdms_lite,super_cdms,cdms_2015}) whose spin is
mostly originated by an unpaired neutron, as well as by the KIMS
experiment\cite{kims,kims_modulation,kims2} which uses $CsI$ and thus
directly probes the contribution to the DAMA effect from WIMP
scatterings off iodine targets. If the WIMP mass is small enough to
assume that the DAMA signal is only due to WIMP scatterings off sodium
the KIMS constraint can be evaded. Moreover, both sodium and iodine in
DAMA have an unpaired proton, so that if the WIMP particle interacts
with ordinary matter predominantly via a spin--dependent coupling
which is suppressed for neutrons it can explain the DAMA effect in
compliance with the bounds from xenon and germanium detectors, whose
constraints are strongly
relaxed\cite{spin_n_suppression,spin_gelmini}. However this scenario
is constrained by droplet detectors (SIMPLE\cite{simple},
COUPP\cite{coupp}) and bubble chambers (PICASSO\cite{picasso},
PICO-2L\cite{pico2l,pico2l_2016},PICO-60\cite{pico60}) which all use
nuclear targets with an unpaired proton (in particular, they all
contain $^{19}F$, while SIMPLE contains also $^{35}Cl$ and $^{37}Cl$
and COUPP and PICO-60 use also $^{127}I$).  As a consequence, this
class of experiments rules out a DAMA explanation in terms of WIMPs
with a spin--dependent coupling to protons when standard assumptions are
made on the WIMP local density and velocity distribution in our
Galaxy\cite{spin_gelmini,pico2l}.

In Ref.\cite{noi_idm_spin} the alternative approach of IDM was
proposed to reconcile DAMA to fluorine detectors. In this class of
models a DM particle $\chi_1$ of mass $m_{\chi_1}=m_{\chi}$ interacts
with atomic nuclei exclusively by up--scattering to a second heavier
state $\chi_2$ with mass $m_{\chi_2}=m_{\chi}+\delta$. A peculiar
feature of IDM is that there is a minimal WIMP incoming speed in the
lab frame matching the kinematic threshold for inelastic upscatters
and given by:

\begin{equation}
v_{min}^{*}=\sqrt{\frac{2\delta}{\mu_{\chi N}}},
\label{eq:vstar}
\end{equation}

\noindent with $\mu_{\chi N}$ the WIMP--nucleus reduced
mass. This quantity corresponds to the lower bound of the minimal
velocity $v_{min}$ (also defined in the lab frame) required to deposit
a given recoil energy $E_R$ in the detector:

\begin{equation}
v_{min}=\frac{1}{\sqrt{2 m_N E_R}}\left | \frac{m_NE_R}{\mu_{\chi N}}+\delta \right |,
\label{eq:vmin}
\end{equation}

\noindent with $m_N$ the nuclear mass. In particular, indicating with
$v_{min}^{*Na}$ and $v_{min}^{*F}$ the values of $v_{min}^*$ for
sodium and fluorine, and with $v_{cut}$ the result of the boost in the
lab rest frame of some maximal speed value beyond which the WIMP
velocity distribution $f(v)$ in the galactic rest frame vanishes
(typically $v_{cut}$ is identified with the WIMP escape velocity
$v_{esc}$), constraints from droplet detectors and bubble chambers can
be evaded when the WIMP mass $m_{\chi}$ and the mass gap $\delta$ are
chosen in such a way that the hierarchy:

\begin{equation}
v_{min}^{*Na}<v_{cut}^{lab}<v_{min}^{*F},
\label{eq:hierarchy}
\end{equation}

\noindent is achieved, since in such case WIMP scatterings off
fluorine turn kinematically impossible while those off sodium can
still serve as an explanation to the DAMA effect. Clearly, this
mechanism rests on the trivial observation that the velocity
$v_{min}^*$ for fluorine is larger than that for sodium.

In Ref. \cite{noi_idm_spin} only a standard spin--dependent coupling
between the WIMP and the nucleon ${\cal N}=p,n$, ${\cal L}_{int}\ni
c^p\vec{S}_{\chi}\cdot \vec{S}_p+c^n\vec{S}_{\chi}\cdot \vec{S}_n$
with $c_n\ll c_p$ was considered (the exact value of $c_n/c_p$ must be
tuned to a different small number depending on the spin--dependent
form factor that is used \cite{noi_idm_spin}). In the present paper we
keep the condition $c_n\ll c_p$ and extend that analysis to the most
general class of spin--dependent interactions\cite{noi_eft_spin} by
making use of the interaction Hamiltonian which descends from
non--relativistic EFT\cite{haxton1,haxton2}:

\begin{eqnarray}
{\bf\mathcal{H}}&=& \sum_{\tau=0,1} \sum_{k=1}^{15}
\frac{c_k^{\tau}}{\Lambda^2} \mathcal{O}_{k} \, t^{\tau} ,
\label{eq:H}
\end{eqnarray}

\noindent where $t^0=1$, $t^1=\tau_3$ denote the the $2\times2$
identity and third Pauli matrix in isospin space, respectively, the
non--dimensional isoscalar and isovector coupling constants $c^0_k$
and $c^{1}_k$ are related to those to protons and neutrons $c^{p}_k$
and $c^{n}_k$ by $c^{p}_k=(c^{0}_k+c^{1}_k)/2$ and
$c^{n}_k=(c^{0}_k-c^{1}_k)/2$, $\Lambda$ is a dimension--1 cut--off
scale and the operators ${\cal O}_i$ are for instance listed in
Equations (12) and (13) of \cite{haxton2}: the corresponding
interaction operators with an exclusively spin--dependent nuclear
response function (i.e. depending only on either the function
$W^{\tau\tau^{\prime}}_{\Sigma^{\prime}}$ or
$W^{\tau\tau^{\prime}}_{\Sigma^{\prime\prime}}$ in the notation of
\cite{noi_idm_spin}) are summarized in Table \ref{table:ref_nref},
where we follow the naming convention $\bar{\chi}_1\Gamma^a\chi_2
\bar{{\cal N}} \Gamma^{b} {\cal N}$+h.c.$\rightarrow {\cal O}_i^{ab}$
with $\Gamma^S=1$, $\Gamma^P=\gamma_5$, $\Gamma^V=\gamma^{\mu}$,
$\Gamma^A=\gamma^{\mu}\gamma_5$, $\Gamma^T$=$\sigma^{\mu\nu}$=$i/2
[\gamma^{\mu},\gamma^{\nu}]$,
$\Gamma^{T'}$=$\sigma^{\mu\nu}\gamma_5$. Following this convention the
standard spin--dependent coupling corresponds to
$\bar{\chi}_1\gamma^{\mu}\gamma_5\chi_2 \bar{{\cal
    N}}\gamma_{\mu}\gamma_5 {\cal N}$ = $\bar{\chi}_1\Gamma^A\chi_2
\bar{{\cal N}} \Gamma^{A} {\cal N}$+h.c.$\rightarrow
\vec{S}_{\chi}\cdot \vec{S}_{{\cal N}}\equiv {\cal O}_4^{AA}$ with
$\Gamma^A=\gamma^{\mu}\gamma_5$ and ${\cal O}_4$=$\vec{S}_{\chi}\cdot
\vec{S}_{{\cal N}}$.

\begin{table}[t]
\tiny
\begin{center}
{\begin{tabular}{@{}|c|c|c|c|c|c|@{}}
\hline
 & Relativistic EFT  &  Non-relativistic limit  & $\sum_i {\cal O}_i$ & cross section scaling\\
\hline\hline
${\cal O}_4^{AA}$ & $\bar{\chi}_1\gamma^{\mu}\gamma^5\chi_2 \bar{{\cal N}}\gamma_{\mu}\gamma^5 {\cal N}$+h.c.  & $-4 \vec{S}_{\chi}\cdot\vec{S}_{\cal N}$ & -4${\cal O}_{4}$ & $W^{\tau\tau^{\prime}}_{\Sigma^{\prime\prime}}(q^2)+W^{\tau\tau^{\prime}}_{\Sigma^{\prime}}(q^2)$ \\
\hline
${\cal O}_9^{VA}$ & $\bar{\chi}_1\gamma^{\mu}\chi_2 \bar{{\cal N}}\gamma_{\mu}\gamma^5 {\cal N}$+h.c.  & $+\frac{2}{m_{\chi}}i\vec{S}_{\chi}\cdot\left (\vec{S}_{\cal N}\times\vec{q} \right )$ & $\simeq2\frac{m_{\cal N}}{m_{\chi}}{\cal O}_9$  & $\simeq q^2W^{\tau\tau^{\prime}}_{\Sigma^{\prime}}(q^2)$ \\
\hline
${\cal O}_9^{TA}$ & $\bar{\chi}_1i\sigma_{\mu\nu}\frac{q^{\nu}}{m_M}\chi_2 \bar{{\cal N}}\gamma^{\mu}\gamma_5 {\cal N}$+h.c.  & $4i  (\frac{\vec{q}}{m_M}\times \vec{S}_{\chi}) \cdot \vec{S}_{\cal N} $ & $4\frac{m_{\cal N}}{m_M}{\cal O}_{9}$ & $q^2 W^{\tau\tau^{\prime}}_{\Sigma^{\prime}}(q^2)$ \\
\hline
${\cal O}_9^{AT}$ & $\bar{\chi}_1\gamma^{\mu}\gamma_5\chi_2 \bar{{\cal N}}i\sigma_{\mu\nu}\frac{q^{\nu}}{m_M} {\cal N}$+h.c.  & $4i\vec{S}_{\chi}\cdot (\frac{\vec{q}}{m_M}\times \vec{S}_{\cal N})$ & $-4\frac{m_{\cal N}}{m_M}{\cal O}_{9}$ & $q^2 W^{\tau\tau^{\prime}}_{\Sigma^{\prime}}(q^2)$ \\
\hline
${\cal O}_{10}^{SP}$ & $i\bar{\chi}_1\chi_2 \bar{{\cal N}}\gamma^5 {\cal N}$+h.c.  & $i\frac{\vec{q}}{m_{\cal N}} \cdot \vec{S}_{\cal N}$+h.c. & ${\cal O}_{10}$ & $q^2 W^{\tau\tau^{\prime}}_{\Sigma^{\prime\prime}}(q^2)$ \\
\hline
${\cal O}_6^{PP}$ & $\bar{\chi}_1\gamma_5\chi_2 \bar{{\cal N}}\gamma^5 {\cal N}$+h.c.  & $-\frac{\vec{q}}{m_{\chi}} \cdot \vec{S}_{\chi}\frac{\vec{q}}{m_{\cal N}} \cdot \vec{S}_{\cal N}$ & $-\frac{m_{\cal N}}{m_{\chi}}{\cal O}_{6}$ & $q^4 W^{\tau\tau^{\prime}}_{\Sigma^{\prime\prime}}(q^2)$ \\
\hline
${\cal O}_6^{T'T'}$ & $\bar{\chi}_1i\sigma^{\mu\alpha}\frac{q_{\alpha}}{m_M}\gamma_5\chi_2 \bar{{\cal N}}i\sigma_{\mu\beta}\frac{q^{\beta}}{m_M}\gamma_5 {\cal N}$+h.c.  &$4\frac{\vec{q}}{m_M}\cdot\vec{S}_{\chi} \frac{\vec{q}}{m_M}\cdot\vec{S}_{\cal N} $ & $4\frac{m_{\cal N}^2}{m^2_M} {\cal O}_6$ & $q^4 W^{\tau\tau^{\prime}}_{\Sigma^{\prime\prime}}(q^2)$  \\
\hline
\end{tabular}}
\caption{Relativistic Effective Field Theories for the nucleon--IDM
  interaction having as a low--energy limit a generalized
  spin--dependent elastic scattering. We do not include models leading
  to an explicit velocity dependence. The quantities $m_{\cal N}$,
  $\vec{S}_{\cal N}$ and $q$ represent the nucleon mass, the nucleon
  spin and the transferred momentum, respectively. Some of the
  interaction terms in the second column contain an arbitrary scale
  $m_M$ to ensure correct dimensionality: in our analysis we will fix
  it to the nucleon mass $m_{\cal N}$. Adapted from Table 1 of
  Ref. \cite{noi_eft_spin}.
\label{table:ref_nref}}
\end{center}
\end{table}

\section{Relic abundance}
\label{sec:relic}

A particularly attractive feature of the WIMP scenario is that the
interactions that keep the DM and Standard Model particles in thermal
equilibrium in the early Universe and that drive the decoupling
process fixing the DM relic abundance can be correlated to those which
may allow their direct or indirect detection.  In particular, in
Section \ref{sec:bottom_up} we will adopt the ``bottom--up'' procedure
of fixing the effective Hamiltonian of Eq. (\ref{eq:H}) by requiring
that it explains the observed DAMA modulation signal, and using it to
calculate the relic abundance of particles $\chi_1$ and $\chi_2$. The
total density of the two $\chi_1$, $\chi_2$ states and the ratio
between their abundances depend on the annihilation process $\chi_1
\chi_2\rightarrow \bar{f} f$ of the DM particles to Standard Model
fermions $f$ and the decay amplitude for the process
$\chi_2\rightarrow \chi_1$. In order to calculate both processes we
make the assumption that the physics of annihilation and decay is
dominated at the quark level by the same 6--dimension operator whose
non--relativistic limit drives direct detection. As summarized in
Table \ref{table:ref_nref} the operators ${\cal O}_k$ of the effective
Hamiltonian (\ref{eq:H}) can be interpreted as the non--relativistic
limits of the interaction terms:

\begin{equation}
   {\cal L} = \frac{c_p}{\Lambda^2}{\cal
       O}^{XY}_k=\frac{c_p}{\Lambda^2} \bar{\chi}\Gamma^X\chi
       \bar{N}\Gamma^Y N,
   \label{eq:lagrangian_nucleon}
\end{equation}
\noindent with XY=AA,VA,TA,AT,SP,PP,T'T'. The above interaction is
generated by an analogous WIMP--quark Lagrangian:

\begin{equation}
   {\cal L} = \sum_q \frac{c_q}{\Lambda^2}\bar{\chi}\Gamma^X\chi \bar{q}\Gamma^Y q,
   \label{eq:lagrangian_quark}
   \end{equation}
\noindent (the relation between the couplings $c_p$ and $c_q$ is
discussed in Appendix \ref{app:matrix_elements}). As far as the decay
process is concerned, the Lagrangian of Eq.(\ref{eq:lagrangian_quark})
induces the decay $\chi_2\rightarrow \chi_1 \gamma\gamma$ through a
quark loop. The scale of the process is of order $\delta~{\cal
  O}(keV)$, and Chiral Perturbation Theory (ChPT) can be used to
evaluate it. For the case of the ${\cal O}_4^{AA}$ operator it is
explicitly given by (see Eq.(3) of Ref.\cite{dienes}):

\begin{equation}
\Gamma_{\gamma\gamma}=\left (\frac{16 \pi^2
    f_{\pi}^2}{m_{\pi}^2}\right )^2\frac{\alpha_{em}^2\delta^9}{512(315
  \pi^9)f_{\pi}^4}\frac{c_4^2}{\Lambda^4},
\label{eq:decay_amplitude_4AA}
\end{equation}

\noindent with $m_{\pi}\simeq$ 140 MeV, $f_{\pi}\simeq$ 93
MeV. Substituting numbers one gets:

\begin{equation}
\Gamma_{\gamma\gamma}=7.2\times 10^{-56}\left(\frac{\delta}{\mbox{10
      keV}} \right )^9 \left (\frac{\mbox{10 GeV}}{\tilde{\Lambda}}
  \right )^4 \mbox{GeV},
\end{equation}

\noindent with $1/\tilde{\Lambda}^2=c_4/\Lambda^2$. This implies that
for the ranges of parameters we will find in Section
\ref{sec:analysis} ($\tilde{\Lambda}\gsim$ 100 GeV, 10 keV$\le
\delta\le$ 20 keV) the lifetime of the $\chi_2$ particle is far larger
than the age of the Universe (corresponding in natural units to
1.5$\times$10$^{-42}$ GeV$^{-1}$) and for practical purposes $\chi_2$
is stable. The same happens for the decay amplitudes calculated for
the other models listed in Table \ref{table:ref_nref}. In particular,
the decay amplitude of model ${\cal O}_9^{VA}$ turns out to be the
same as that of ${\cal O}_4^{AA}$, while models with tensor currents,
i.e. ${\cal O}_9^{TA}$, ${\cal O}_9^{AT}$ and ${\cal O}_6^{T'T'}$ lead
to vanishing decay amplitudes at leading order in ChPT; finally, the
decay amplitudes for models ${\cal O}_{10}^{SP}$ and ${\cal O}_6^{PP}$
turn out to be enhanced by a factor $(f_{\pi}/\delta)^2$ and
suppressed by a factor $f_{\pi}/m_{\chi}$ compared to
Eq. (\ref{eq:decay_amplitude_4AA}), respectively, with no practical
phenomenological differences with the other models. The bottom line is
that for all the models of Table \ref{table:ref_nref}, if no other
coupling is assumed for the DM particle, our Galaxy is formed by a
mixture of $\chi_1$ and $\chi_2$, which, presumably, are in equal
parts and have the same velocity distribution, and direct detection
experiments are both sensitive to the upscatters of $\chi_1$'s and the
downscatters of $\chi_2$'s.

This is a problem in the scenario outlined in Section
\ref{sec:spin_idm_scenario}, where IDM upscatters off fluorine must be
kinematically suppressed in order to comply with the constraints from
bubble chambers and droplet detectors, and which requires $v_{min}^* >
v_{esc}$, because $v_{min}^*$ vanishes when $\delta<0$, i.e. for the
downscatters of the $\chi_2$ particles the condition
(\ref{eq:hierarchy}) is never achieved. A simple numerical check using
the compatibility factors introduced in Section \ref{sec:analysis}
confirms that the downscatters of $\chi_2$ rule out such scenario
experimentally. A possible way out is adding to the theory a
tree--level coupling to neutrinos. In such case, assuming for
instance:

\begin{equation}
  {\cal L}_{int,\nu}=\frac{c_{\nu}}{\Lambda^2}(\bar{\chi}_2\gamma_{\mu}\gamma_5 \chi_1)(\bar{\nu}\gamma^{\mu}\gamma_5 \nu),
  \label{eq:l_neutrino}
\end{equation}

\noindent the decay amplitude to standard left--handed neutrinos turns
out to be:

\begin{equation}
\Gamma_{\nu\nu}=\frac{11\delta^5}{768 \pi^4}\frac{1}{\tilde{\Lambda}^4},
\end{equation}

\noindent i.e, numerically:

\begin{equation}
\Gamma_{\nu\nu}=1.5\times 10^{-33}\left(\frac{\delta}{\mbox{10
      keV}} \right )^5 \left (\frac{\mbox{10 GeV}}{\tilde{\Lambda}}
  \right )^4 \mbox{GeV},
\end{equation}

\noindent (where $\tilde{\Lambda}^2=\Lambda^2/|c_{\nu}|$), so that the
$\chi_2$ particle can safely decay after decoupling from the plasma
and the halo of our Galaxy is only made of $\chi_1$ states. Moreover,
we notice that Eq.(\ref{eq:l_neutrino}) allows for a fast decay of
$\chi_2$ already for $|c_{\nu}|<<|c_q|$; in such a case the DM thermal
decoupling from the plasma and its relic density are only driven by
the couplings to quarks responsible of direct detection, and the
correlation between direct detection rates and the relic density is
maintained.

In the following we will assume the presence of the coupling of
Eq.(\ref{eq:l_neutrino}) with $|c_{\nu}|<<|c_q|$ or an analogous
mechanism that depletes the $\chi_2$ particles without affecting the
relic density. In such case the DM particle relic abundance is given
by a standard expression inversely proportional to the quantity
$\tilde{<\sigma v>}=a+b/(2 x_f)$, where $<\sigma v>=a+b/x$ is the
thermal average of the coannihilation cross section at temperature
$T=m_{\chi}/x$, and $x_f\simeq 20$ corresponds to the freeze-out
temperature (in Appendix \ref{app:sigmav} the coefficients $a$ and $b$
for the different models listed in Table \ref{table:ref_nref} are
provided). For minimality, we will only assume a coupling of the DM
particles to the quarks that drive spin--dependent direct detection,
i.e. $f$=$u$, $d$, $s$ for models ${\cal O}^{AA}_4$, ${\cal
  O}^{VA}_9$, ${\cal O}^{TA}_9$, ${\cal O}^{SP}_{10}$, ${\cal
  O}^{PP}_6$ and $f$=$u$, $d$ for models ${\cal O}^{AT}_{9}$ and
${\cal O}^{TT}_{6}$ (see Appendix \ref{app:matrix_elements}).
Combined with the condition $|c_{\nu}|<<|c_q|$, this implies that all
the quantitative values of the relic abundance discussed in Section
\ref{sec:analysis} will be be upper bounds. Indeed, we will see in
Section \ref{sec:analysis} that this represents the most favorable
choice to explain the DAMA effect and get at the same time an
acceptable relic abundance through thermal decoupling in a consistent
way. In our numerical analysis we will not make any further
assumptions on the $c_q$ parameters, besides imposing perturbativity
($|c_q|<4 \pi$) and $c_n\ll c_p$, and we will neglect the effect of
the running of the couplings\cite{cq_running}\footnote{In some cases
  (such as model VA in Table
  \ref{table:ref_nref}\cite{running_deramo1, running_deramo2,
    running_crivellin}) the ultraviolet completion of our scenario
  requires some level of tuning at high energy, since the running of
  the couplings generates a non--vanishing SI interaction at low
  energy.}. We conclude this section by pointing out that, obviously,
the scenario outlined above does not produce any indirect detection
signal.

\section{The bottom--up approach}
\label{sec:bottom_up}

In the present paper the data of DAMA and of the other direct
detection experiments will be used to determine estimations or upper
bounds of the halo functions $\tilde{\eta}_0$, $\tilde{\eta}_1$
defined as:

\begin{equation}
  \tilde{\eta}_{0,1}(v_{min})=\frac{\rho_{\chi}}{m_{\chi}}\sigma_{ref}\eta_{0,1}(v_{min})
  =\frac{\xi\rho_{loc}}{m_{\chi}}\sigma_{ref}\eta_{0,1}(v_{min}),
\label{eq:eta_tilde}  
\end{equation}

\noindent with $\rho_{loc}$=0.3 GeV/cm$^2$ the observed total DM
density in the neighborhood of the Sun, $\rho_{\chi}$=$\xi \rho_{loc}$
the corresponding local density of the $\chi_1$ particles
(contributing in general a fraction $\xi$ of $\rho_{loc}$),
$\sigma_{ref}$ represents a reference WIMP--nucleon cross section that
we take as:

\begin{equation}
\sigma_{ref}=\frac{(c^p_k)^2}{\Lambda^4} \mu^2_{\chi {\cal N}}/\pi,
  \label{eq:sigma_ref}
\end{equation}
\noindent for a given operator ${\cal O}_k$, while:

\begin{eqnarray}
\eta_0(v_{min},t)&=&\int_{v_{min}}^{\infty} \frac{f(v,t)}{v}\,dv,\label{eq:eta0_t}\\
\eta_0(v_{min})&=&\frac{1}{T}\int_0^T \eta_0(v_{min},t)\,dt,\label{eq:eta0}\\
  \eta_1(v_{min})&=&\frac{2}{T} \int_0^T
  \cos (\omega (t-t_0)) \eta_0(v_{min},t)\,dt,\label{eq:eta1}
\end{eqnarray}  

\noindent where the time--dependence of the velocity distribution
$f(v,t)$ descends from the rotation of the Earth around the Sun with
$\omega=2\pi/(\mbox{365 days})$ and phase $t_0\simeq$2 June, and is due
to the boost from the galactic to the Earth rest frame, while
$v_{min}$ is given by Eq.(\ref{eq:vmin}). The halo functions of
Eqs. (\ref{eq:eta0},\ref{eq:eta1}) are defined in such a way that, for a
given experimental setup, the corresponding expected direct detection
signal $S^{[E_1,E_2]}(t)$ in the observed energy interval $[E_1,E_2]$
is given by\cite{gondolo_eta1,gondolo_eta2,gondolo_eta3}:

\begin{eqnarray}
S^{[E_1,E_2]}(t)&=&S_0^{[E_1,E_2]}+S_1^{[E_1,E_2]}\cos(\omega(t-t_0),\nonumber\\
S_{0,1}^{[E_1,E_2]}&=&\int_0^{\infty} {\cal R}^{[E_1,E_2]}(v_{min})\tilde{\eta}_{0,1}(v_{min})\label{eq:s01}\,d v_{min}.
\end{eqnarray}

\noindent In particular, the response functions ${\cal
  R}^{[E_1,E_2]}(v_{min})$ are supposed to be known since they depend
only on the properties of the detector, while the functions
$\tilde{\eta}_{0,1}$ depend both on particle physics (through
$\sigma_{ref}$) and on astrophysics (through $\rho_{\chi}$ and $f(v)$)
and factorize all our ignorance about the process. Explicit
expressions for the response functions ${\cal R}$ for the different
experiments considered in this paper are given in
\cite{noi1,noi_eft_spin}.

If no assumptions are made about $f(v)$ the two halo functions
$\tilde{\eta}_0$ and $\tilde{\eta}_1$ are subject to the very general
conditions:

\begin{eqnarray}
&\tilde{\eta}_0(v_{min,2})\le\tilde{\eta}_0(v_{min,1})  & \mbox{if $v_{min,2}> v_{min,1}$},\nonumber\\
& \tilde{\eta}_1\le\tilde{\eta}_0  & \mbox{at the same $v_{min}$},\nonumber\\
& \tilde{\eta}_{0,1}(v_{min} \ge v_{s})=0. &
\label{eq:eta_conditions}
\end{eqnarray}

The first condition descends from the definition (\ref{eq:eta0_t}),
that implies that $\tilde{\eta}_0(v_{min})$ is a decreasing function
of $v_{min}$. The second is a consequence of the fact that
$\tilde{\eta}_1$ is the modulated part of
$\tilde{\eta}_0$\footnote{The use of representation theorems for
  distribution functions of the type (\ref{eq:s01}) may allows to
  restrict further the range of the ratio
  $\tilde{\eta}_1/\tilde{\eta}_0$ in a statistical way even without
  specifying the velocity distribution
  \cite{scopel_gondolo_unmodulated}.}. The last condition with $v_s$=
$v_{esc}$ reflects the requirement that the WIMPs are gravitationally
bound to our Galaxy. We will adopt this choice in our Maxwellian
analysis and, in this case, take as the escape velocity of WIMPs in
the lab rest frame $v_{esc}$=$v_{esc}^{Galaxy}$+$v_{Sun}$, with
$v_{esc}^{Galaxy}$=550 km/sec the escape velocity in the galactic rest
frame and $v_{Sun}$=232 km/sec the velocity of the Solar system with
respect to the WIMP halo. On the other hand, in our halo--independent
analysis $v_s$ will represent the maximal value of the $v_{min}$ range
corresponding to the DAMA excess. Not that in this case $v_s$ will
depend on the model parameters $m_{\chi}$ and $\delta$: as explained
below it will be adopted in order to get the largest $\eta_1(v_{min})$
which does not vanish in the signal range.

Intuitively, the measurement of a signal in a direct detection
experiment such as that of DAMA should allow to determine the
WIMP--nucleon cross section $\sigma_{ref}$ and in this way the EFT
cut--off scale $\Lambda$, shedding light on the scale of the new
physics involved in the process. Notice, however, that
Eq.(\ref{eq:eta_tilde}) implies that direct detection allows
to get access to the product $\xi\sigma_{ref}$ rather then
$\sigma_{ref}$. As a consequence, once an experimental estimate
$(\xi\sigma_{ref})_{exp}$ is obtained from the data, using
Eq.(\ref{eq:sigma_ref}) any direct detection experiment is sensitive
to the effective scale:

\begin{equation}
\tilde{\Lambda}_{p,exp}=\frac{\Lambda}{\xi^{1/4}|c_p|^{1/2}}=\left(\frac{\mu_{\chi{\cal
      N}}^2}{\pi (\xi\sigma_{ref})_{exp}}\right)^{\frac{1}{4}},
  \label{eq:lambda_tilde_p}
\end{equation}

\noindent rather than directly to the physical cut--off scale
$\Lambda$.

The firs step to measure $ (\xi\sigma_{ref})_{exp}$ is to get
experimental estimates of the halo functions $\tilde{\eta}_0$ and/or
$\tilde{\eta}_1$ from the data. In particular, given an experiment
with detected count rate or modulation amplitude $N_{exp}$ in the
energy interval $E_1<E<E_2$ the combination\cite{gondolo_eta1}:

\begin{equation}
  <\bar{\tilde{\eta}}_{0,1}>=\frac{\int_{v_{min}^*}^{\infty} d v_{min} \tilde{\eta}_{0,1}(v_{min}) {\cal R}_{[E_1,E_2]}(v_{min})}
  {\int_{v_{min}^*}^{\infty} d v_{min} {\cal R}_{[E_1,E_2]}(v_{min})}=
  \frac{N_{exp}}{\int_{v_{min}^*}^{\infty} d v_{min} {\cal R}_{[E_1,E_2]}(v_{min})},
\label{eq:eta_bar_vt}
\end{equation}

\noindent can be interpreted as an average of the halo function
$\tilde{\eta}_{0,1}(v_{min})$ in an interval
$v_{min,1}<v_{min}<v_{min,2}$. The $v_{min}$ interval is defined as
the one where the response function ${\cal R}$ is ``sizeably''
different from zero, including the smearing effect of energy
resolution.

Getting $(\xi\sigma_{ref})_{exp}$ from the
$<\bar{\tilde{\eta}}_{0,1}>$'s requires to fix the velocity
distribution $f(v)$. In this way the function $\eta_1(v_{min})$ is
known and $\xi\sigma_{ref}$, which is just a normalization factor of
the ensuing $\tilde{\eta}_1(v_{min})$, can be directly fitted from the
data.  If, on the other hand, a halo--independent approach is assumed,
since $\int f(\vec{v}) d^3 v$=1 and $\tilde{\eta}_1\le\tilde{\eta}_0$
the function $\eta_1(v_{min})$ can be at least maximized\cite{noi2} by
the choice $\tilde{\eta}_1$ =$\tilde{\eta}_0$ and
$f(\vec{v})=\delta(v_s-v_{min})$, with $v_s$ the maximal value of the
$v_{min}$ range corresponding to the DAMA excess. This corresponds to
the largest $\eta_1(v_{min})$ which does not vanish in the signal
range, and allows to get a lower bound on
$(\xi\sigma_{ref})_{exp}$. In this case:

\begin{equation}
\tilde{\eta}_1^{max}(v_{min})=\tilde{\eta}_0^{max}(v_{min})=\frac{\bar{\tilde{\eta}}_{1,fit}^{DAMA}}{v_s}
\theta(v_s-v_{min}),
\label{eq:eta_max_vs}
\end{equation}

\noindent where the constant value $\bar{\tilde{\eta}}_{1,fit}^{DAMA}$
can be fitted from the DAMA observed modulation amplitudes in a
straightforward way\footnote{Indeed, due to the large error bars such
  flat functional form for the $\tilde{\eta}_1^{max}(v_{min})$
  function is in general not incompatible with the DAMA experimental
  data.}. Plugging Eqs.(\ref{eq:eta_max_vs}), (\ref{eq:sigma_ref}) and
(\ref{eq:eta_tilde}) in Eq. (\ref{eq:lambda_tilde_p}) one then gets
for the experimental estimate $\tilde{\Lambda}_{p,exp}$ of
$\tilde{\Lambda}_p$ the upper bound:

\begin{equation}
  \tilde{\Lambda}_{p,exp}<\tilde{\Lambda}_{p,exp}^{max}=\left(\frac{ \rho_{loc}
    \mu_{\chi {\cal N}}^2}{\pi \bar{\tilde{\eta}}_{fit}^{DAMA}
    m_{\chi} v_s} \right )^{\frac{1}{4}}.
\label{eq:lambda_tilde_p_halo_indep}
\end{equation}

Notice that the halo--independent procedure allows to determine only a
lower bound on $\sigma_{ref}$, and so an upper bound on
$\tilde{\Lambda}_p$.

The quantity above, which is obtained analyzing experimental data,
must be compared to the corresponding theoretical expectation for a
given model ${\cal O}^{XY}_k$ and choice of the physical cut--off
scale $\Lambda$ and of the couplings $c_q$ (related to $c_p$ through
Eq.(\ref{eq:couplings_qp})).  A crucial observation here is that,
whenever the DM particle calculated relic density $\Omega$ turns out
to be smaller than the observed value $\Omega_0$, $\rho_{\chi}$ must
be rescaled with respect to the observed local DM density $\rho_{loc}$
by the factor $\xi\equiv \Omega/\Omega_0$. To get an expression of the
rescaling factor $\xi$ we can express the coannihilation cross section
in the form:

\begin{equation}
  \tilde{<\sigma v>}=\frac{1}{\Lambda^4}\sum_q c_q^2 \tilde{<\sigma>}_q,\,\,\,\,
  \tilde{<\sigma>}_q= N_q\left [(a)_q+\frac{(b)_q}{2 x_f}\right ],
\label{eq:sigmav1}
\end{equation}  

\noindent with $(a)_q$ and $(b)_q$ given in Appendix \ref{app:sigmav}.
Indicating with $\tilde{<\sigma v>}_0=2\times 10^{-9}$ GeV$^{-2}$ the
value corresponding to the observed relic density, the rescaling
coefficient $\xi$ is then given by:

\begin{equation}
\xi = \frac{\tilde{<\sigma v>}_0}{\tilde{<\sigma
    v>}}=\frac{\Lambda^4 \tilde{<\sigma v>}_0}{\sum_q c_q^2
  \tilde{<\sigma v>}_q}.
\label{eq:rescaling_factor}
\end{equation}
\noindent In this way the theoretical expectation
$\tilde{\Lambda}_{p,th}$ of the effective scale turns out to be:

\begin{equation}
\tilde{\Lambda}_{p,th}=\left(\frac{\sum_q c_q^2 \tilde{<\sigma
    v>_q}}{c_p^2 \tilde{<\sigma
    v>_0}}\right)^{\frac{1}{4}}=
\left(\frac{\sum_q r_q^2 \tilde{<\sigma
    v>_q}}{(c_p/c_u)^2 \tilde{<\sigma
    v>_0}}\right)^{\frac{1}{4}}=
\tilde{\Lambda}_{p,\Omega_0}.
\label{eq:lambda_tilde_p_th}
\end{equation}

\begin{figure}
\begin{center}
\includegraphics[width=0.49\columnwidth, bb=73 193 513 636]{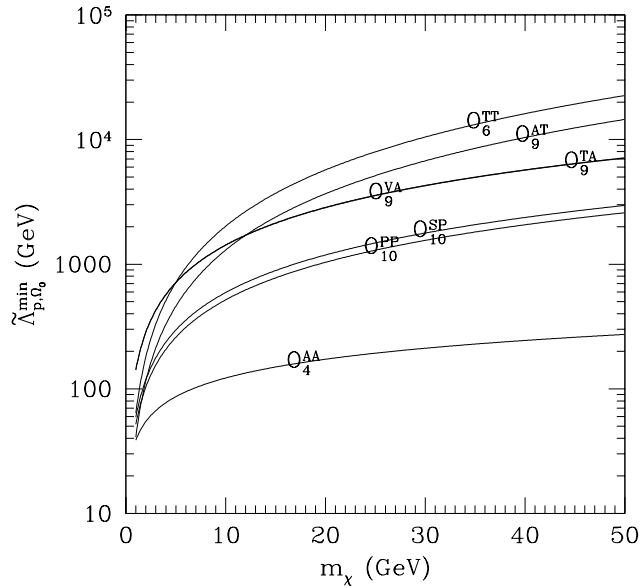}
\end{center}
\caption{Theoretical expectations of
  $\tilde{\Lambda}_{p,\Omega_0}^{min}$ (defined in
  Eq.(\ref{eq:lambda_tilde_p_th})), minimized as a function of the
  coupling ratios $r_q$ and for $c_n$=0, plotted as a function of the
  WIMP mass $m_{\chi}$ for the models of Table \ref{table:ref_nref},
  as indicated by the caption on each curve.}
\label{fig:lambda_p_tilde_min}
\end{figure}

\noindent 
Notice that in the expression above the dependence on the cut-off
scale $\Lambda$ cancels out. Moreover, the second expression in
parenthesis shows explicitly that $\tilde{\Lambda}_{p,th}$ depends
only on the coupling ratios $r_q\equiv c_q/c_u$ and not on the actual
values $c_q$. Finally, $\tilde{\Lambda}_{p,th}$ is equal to the
quantity $\tilde{\Lambda}_{p,\Omega_0}$ that yields $\tilde{<\sigma
  v>}=\tilde{<\sigma v>}_0$, i.e. it corresponds to the observed relic
abundance $\Omega_0$. This means that a direct detection experiment
has in principle no direct access to the physical scale $\Lambda$ and
is bound to measure $\tilde{\Lambda}_p=\tilde{\Lambda}_{p,\Omega_0}$
even if $\Omega\ll\Omega_0$\footnote{This is at variance with indirect
  detection where signals scale with $\rho_{loc}^2$ and are maximal
  when $\Omega=\Omega_0$\cite{subdominant_indirect}}, irrespective of
the actual size of the couplings.  Actually, the requirement
$\tilde{\Lambda}_{p,exp}$=$\tilde{\Lambda}_{p,th}$ combined with
Eqs.(\ref{eq:lambda_tilde_p_halo_indep}) and
(\ref{eq:lambda_tilde_p_th}) implies:

\begin{equation}
  \tilde{\Lambda}_{p,exp}^{max}>\tilde{\Lambda}_{p,\Omega_0}
  >\tilde{\Lambda}_{p,\Omega_0}^{min},
  \label{eq:relic_condition}
\end{equation}

\noindent where $\tilde{\Lambda}_{p,\Omega_0}^{min}$ represents the
minimum of $\tilde{\Lambda}_{p}$ yielding $\Omega_0$ when the
parameters $r_q$ are varied at fixed $c_n$=0. In
Fig.(\ref{fig:lambda_p_tilde_min})
$\tilde{\Lambda}_{p,\Omega_0}^{min}$ is plotted as a function of
$m_{\chi}$ for the models of Table \ref{table:ref_nref}.

\begin{figure}
\begin{center}
\includegraphics[width=0.49\columnwidth, bb=73 193 513 636]{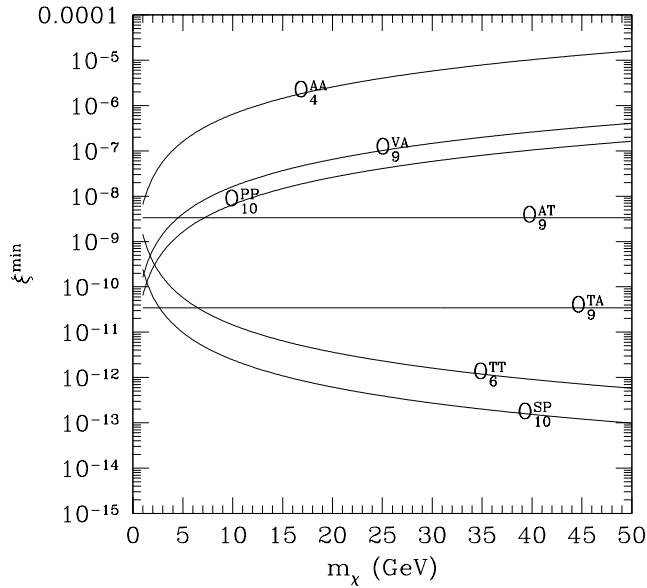}
\end{center}
\caption{The minimal rescaling factor defined in
  Eq. (\ref{eq:rescaling_factor_lower}) and calculated for the same
  $r_q$ coupling ratios used in Fig.\ref{fig:lambda_p_tilde_min} is
  plotted as a function of the WIMP mass $m_{\chi}$ for the models of
  Table \ref{table:ref_nref}, as indicated by the caption on each
  curve.}
\label{fig:csi_min}
\end{figure}

From Eq.(\ref{eq:relic_condition}) one can see that the energy scale
inferred by the experimental data must yield $\Omega>\Omega_0$
(``overclosure'') to be acceptable\footnote{This might seem confusing
  but is simply due to the fact that in a halo--independent approach
  Eq.(\ref{eq:eta_max_vs}) is used to make inferences on
  $\tilde{\Lambda}_p$; on the other hand the energy scale inferred by
  the experimental data is supposed to yield $\Omega=\Omega_0$ for the
  correct choice of $f(v)$.}. On the other hand, the only direct
information on the relic abundance can be obtained by combining the
requirement that the EFT of Eq. (\ref{eq:lagrangian_quark}) is valid
at the scale of the coannihilation process (implying
$\Lambda>E_{\chi_1}+E_{\chi_2}\simeq 2 m_{\chi}$) with perturbativity
($|c_q|<4\pi$). In this way, taking into account the condition
$c_n$=0, the following lower bound on $\xi$ can be obtained:

\begin{equation}
1>\xi>\xi^{min}=\frac{16 m_{\chi}^4 \tilde{<\sigma
    v>}_0}{max|_{|c_q|\le 4\pi,c_n=0}(\sum_q c_q^2 \tilde{<\sigma
    v>}_q)},
\label{eq:rescaling_factor_lower}
\end{equation}

\noindent where in the last expression the denominator is maximized
with the conditions $|c_q|\le 4\pi$ and $c_n=0$. The quantity above is
plotted in Fig.\ref{fig:csi_min} as a function of the WIMP mass
$m_{\chi}$ for the models of Table \ref{table:ref_nref} for
$max(|c_q|)=4\pi$ but using instead the same $r_q$ coupling ratios
employed in Fig.\ref{fig:lambda_p_tilde_min}. From such figure one can
see that $\xi\ll$ 1 for all the effective models analyzed in the
present paper. This has two consequences: (i) indeed, a positive
signal in a direct detection experiment leaves a very large degeneracy
on the actual value of the relic abundance $\Omega$; (ii) the
constraints from EFT validity and perturbativity from
Eq.(\ref{eq:rescaling_factor_lower}) on the allowed parameter space
are not significant.

The condition (\ref{eq:relic_condition}) is the main result of this
Section and will be discussed in the numerical analysis of the pSIDM
scenario of Section \ref{sec:analysis}. It should be pointed out that
it is not simply a requirement to obtain the correct relic abundance
with the same parameters that explain a given experimental excess:
instead, it should be interpreted as a more general condition on the
possibility that the same effective model that explains such excess
can also determine through thermal decoupling the relic abundance,
irrespective on its actual value. It should also be pointed out that
the degeneracy of direct detection signals on the physical cut--off
scale $\Lambda$ discussed in this Section is in principle resolved by
LHC signals and by indirect detection (although the specific case of
the the scenario discussed in the present paper implies no indirect
signals).

\section{Results}
\label{sec:analysis}

In this Section we wish to determine the regions in the
$m_{\chi}$--$\delta$ parameter space where an interpretation of the
DAMA modulation effect in terms of a signal from the WIMPs of the
pSIDM scenario outlined in Section \ref{sec:spin_idm_scenario} is
compatible to constraints from other experiments.  In order to do so
we follow closely the analysis of Ref.\cite{noi_idm_spin} (to which we
refer for further details) that we extend here to the generalized spin
models listed in Table \ref{table:ref_nref}.  Moreover, in the same
parameter space regions we discuss the requirement
(\ref{eq:relic_condition}), i.e. we show when the same effective model
used to interpret direct detection data can be assumed to calculate
the relic density through thermal decoupling in a consistent way.

Using Eq.(\ref{eq:eta_bar_vt}) it is straightforward, for a given
choice of the DM parameters, to obtain estimations
$\bar{\tilde{\eta}}_{1i}^{DAMA}$ of the modulated halo function
$\tilde{\eta}_1(v_{min})$ averaged in appropriately chosen $v_{min,i}$
intervals mapped from the DAMA experimental annual modulation
amplitudes, and upper bounds $\bar{\tilde{\eta}}_{0j,lim}$ in the same
$v_{min}$ intervals on the functions $\tilde{\eta}_0(v_{min})$ from
the data of the experiments that have reported null results. Thanks to
the condition $\tilde{\eta}_1(v_{min})\le\tilde{\eta}_0(v_{min})$ this
allows to constrain the DAMA result. In particular, we calculate the
response functions ${\cal R}$ as explained in Ref.\cite{noi_idm_spin},
with the nuclear form factors from \cite{haxton1,haxton2} and fix
$c_n/c_p$=-0.03\cite{noi_idm_spin}, which is the value that minimizes
the response on xenon using the form factor of
Ref. \cite{haxton1,haxton2} (this is equivalent to taking $c_n$=0 as
far as Sections \ref{sec:relic} and \ref{sec:bottom_up} are
concerned).

Quantitatively, for a given choice of the WIMP mass $m_{\chi}$ and of
the mass difference $\delta$ the compatibility between DAMA and all
the other results can be assessed introducing the following
compatibility ratio\cite{noi_idm_spin}:

\begin{equation}
  {\cal D}_{\mbox{Halo indep.}} \equiv \max_{i\in \mbox{signal}}
\left (\frac{\bar{\tilde{\eta}}_{1i}^{DAMA}+\sigma_i}{\min_{j\le i}\bar{\tilde{\eta}}_{0j,lim}} \right ),
\label{eq:compatibility_factor_eta_i}
\end{equation}

\noindent where $\sigma_i$ represents the standard deviation on
$\bar{\tilde{\eta}}_{1i}^{DAMA}$ as estimated from the data, $i\in
\mbox{signal}$ means that the maximum of the ratio in parenthesis is
for $v_{min,i}$ corresponding to the DAMA excess, while, due to the
fact that the function $\tilde{\eta}_0$ is non--decreasing in all
velocity bins $v_{min,i}$, the denominator contains the most
constraining bound on $\tilde{\eta}_0$ for $v_{min,j}\le v_{min,i}$.
The latter minimum includes all available bounds from scintillators,
ionizators and calorimeters (see Appendix \ref{app:exp} of the present
paper, Appendix B of Ref.\cite{noi_eft_spin} an Appendix A of
\cite{noi_idm_spin} for a summary of the experimental inputs used in
our analysis). Specifically, compatibility between DAMA and the
constraints included in the calculation of
Eq.(\ref{eq:compatibility_factor_eta_i}) is ensured if ${\cal
  D}_{\mbox{Halo indep.}}<1$.

The above procedure cannot be applied to bubble chambers and droplet
detectors, which contain different nuclear targets and, being only
sensitive to the energy threshold, do not allow to map the
corresponding bounds to arbitrary velocity bins. In this case,
however, an alternative procedure can be applied, which consists to
derive from the measured DAMA modulation amplitudes a piece--wise
estimation $\tilde{\eta}^{est}_0(v_{min})$ of the smallest
non--increasing halo function that can explain the effect (this
includes the condition $\tilde{\eta}^{est}_0(v_{min}>v_s)$=0, with
$v_s$ the maximal value of the $v_{min}$ range corresponding to the
DAMA excess), and to use it to calculate for each experiment among
$k$=SIMPLE, COUPP, PICO-60, PICASSO and PICO-2L and for each of the
corresponding energy thresholds $E_{th,i}$ the expected number of WIMP
events $N_{k,i}^{expected}$ and compare it to the corresponding 90\%
C.L. upper bound $N_{k,i}^{bound}$ (see Appendix B of
Ref\cite{noi_eft_spin}, Appendix A of Ref. \cite{noi_idm_spin} and
Appendix \ref{app:exp} of the present paper for further details, and
Fig.\ref{fig:c4_eta_vmin} for an explicit example of the
$\tilde{\eta}^{est}$ function determination from the data).  Then a
straightforward generalization of the compatibility factor of
Eq.(\ref{eq:compatibility_factor_eta_i}) is
\footnote{In our numerical analysis we use a refined version of
  Eq.(\ref{eq:compatibility_factor_generilized}), given in Eq. (3.9) of
  \cite{noi_idm_spin}.}:
  
\begin{equation} {\cal D}_{\mbox{Halo indep.}}\rightarrow \max \left ({\cal
  D}_{\mbox{Halo indep.}}, \frac{N_{k,i}^{expected}}{N_{k,i}^{bound}} \right ).
\label{eq:compatibility_factor_generilized}
\end{equation}

In this Section we will also test the compatibility of the DAMA
modulation effect when the WIMP velocity distribution is fixed to a
standard Maxwellian at rest in the galactic rest frame. Specifically,
we adopt the r.m.s. velocity $v_{rms}$=270 km/s, cut it at the escape
velocity $v_{esc}^{Galaxy}$=550 km/sec and boost it to the lab rest
frame through $v_{sun}$=232 km/s. In this case for a fixed value of
the WIMP mass the whole experimental spectrum allows to get directly a
single upper bound, or in case of an excess, a single estimation of
the combination $\xi\sigma_{ref}$. As far as the DAMA modulation
effect is concerned, we estimate an interval
$\sigma_{DAMA}^{min}\le\sigma_{eff}\le\sigma_{DAMA}^{max}$ for the
cross section by minimizing a $\chi$--square of the $\tilde{\eta}_1$
function through the corresponding experimental estimations (in this
case $\xi\sigma_{eff}$ is just a normalization factor of
$\tilde{\eta}_1$), and apply a quality check on the corresponding
prediction for the modulation amplitudes, requiring that the
$p$--value of the minimal $\chi$--square exceeds $p_{min}$=0.05.  In
the case of null experiments we will use the same energy bins used for
the halo--independent approach, and add one last bin containing the
whole experimental range analyzed in the experiment. For each
experiment $k$ and each energy bin $i$ we will then calculate the
corresponding upper bound $\sigma_{ki}^{bound}$.  Then for the
Maxwellian case we adopt the compatibility factor:

\begin{equation} 
{\cal D}_{Maxwellian} 
\equiv \max\left [ \max_{k,i}
\left (\frac{\sigma_{DAMA}^{min}}{\sigma_{ki}^{bound}} \right ),\frac{p_{min}}{p} \right ].
\label{eq:compatibility_factor_maxwellian}
\end{equation}

\noindent Also in this case compatibility between DAMA and constraints
is ensured if ${\cal D}_{Maxwellian}<1$.

\begin{figure}
\begin{center}
\includegraphics[width=0.49\columnwidth, bb=64 197 505 630]{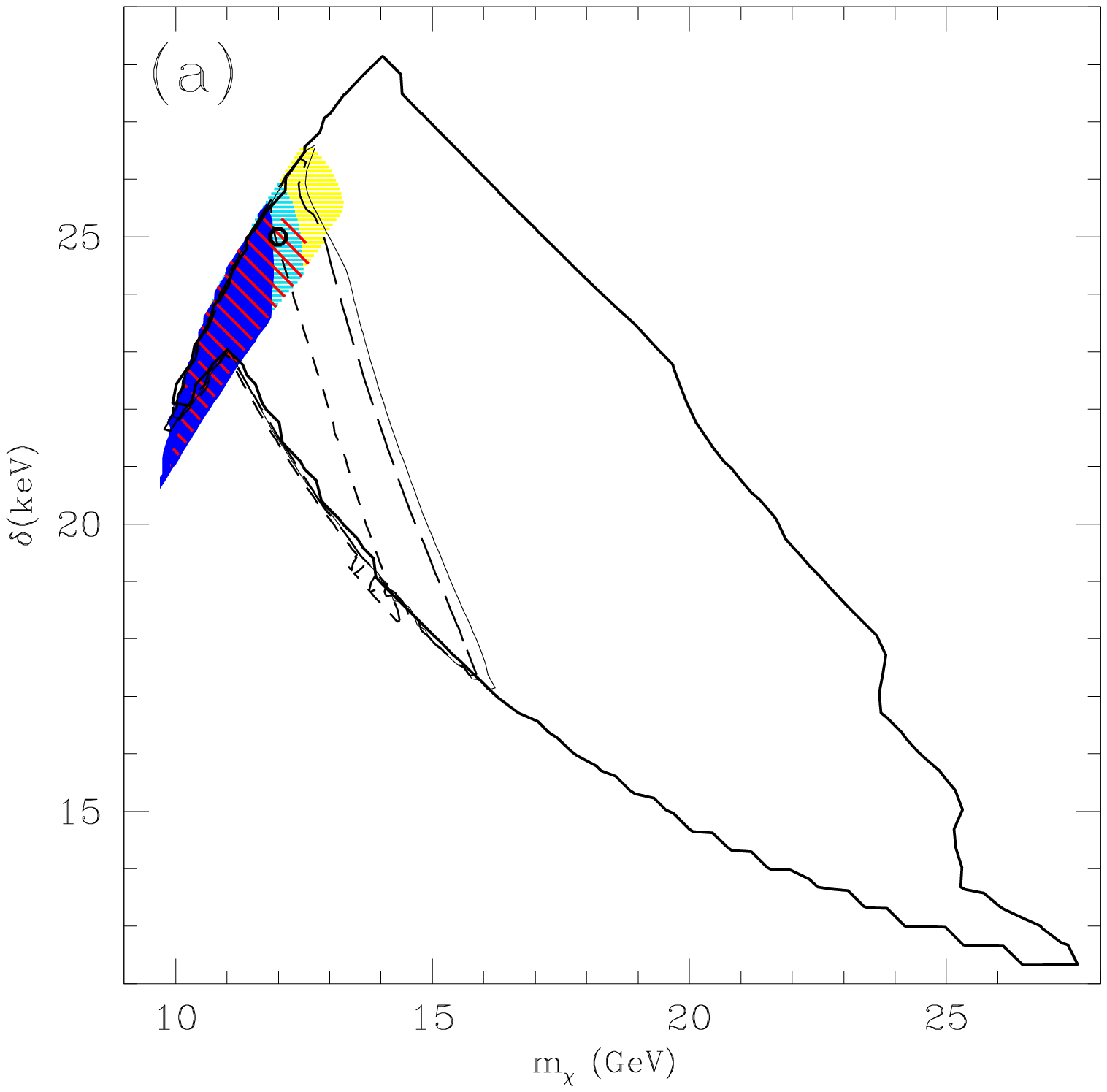} 
\includegraphics[width=0.49\columnwidth, bb=64 197 505 630]{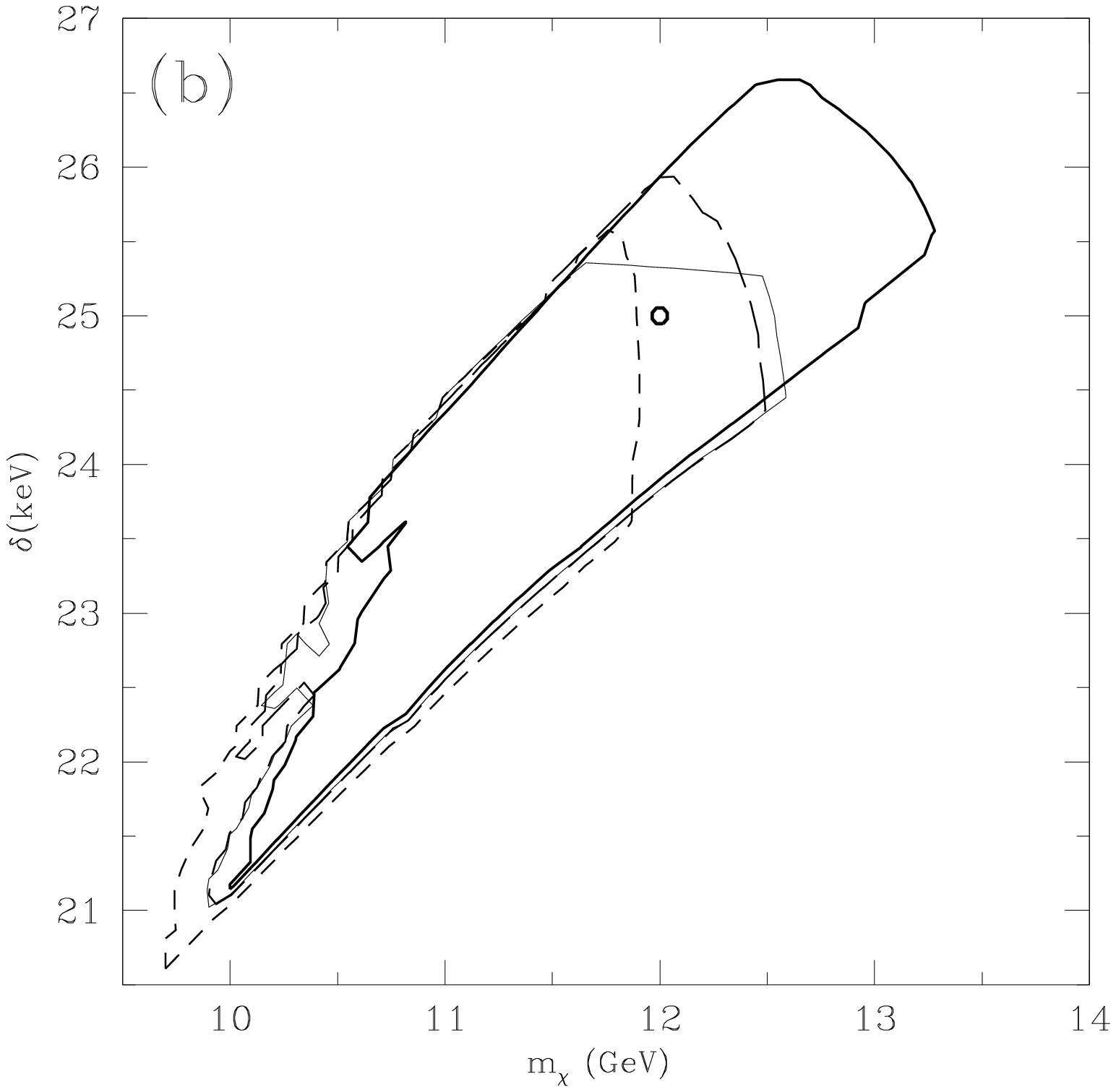} 
\end{center}
\caption{The regions within the contours show the $m_{\chi}$--$\delta$
  parameter space where the compatibility factors defined in
  Eqs.(\ref{eq:compatibility_factor_generilized}) and
  (\ref{eq:compatibility_factor_maxwellian}) are smaller than
  unity. {\bf (a)} Regions enclosed by lines show the result of a
  halo--independent analysis. Thick solid: ${\cal O}_4$ operator; thin
  solid: ${\cal O}_9$; long-dashed: ${\cal O}_6$; short--dashed:
  ${\cal O}_6$. Shaded areas represent the same for a Maxwellian WIMP
  velocity distribution. Light (yellow) shade: ${\cal O}_4$;
  medium--dark (cyan) shade: ${\cal O}_{10}$; slanted (red) dashes:
  ${\cal O}_9$; dark (blue) shade: ${\cal O}_6$.  {\bf (B)} A zoom--up
  of Maxwellian regions: the line code is the same used in plot (a)
  for the halo--independent case. In both plots the open circle
  represents the benchmark point $m_{\chi}$= 12 GeV, $\delta$=25 keV
  discussed in detail in Fig. \ref{fig:c4_eta_vmin}.}
\label{fig:compatibility_all}
\end{figure}

The result of a systematic scan of the compatibility factors ${\cal
  D}_{\mbox{Halo indep.}}$ and ${\cal D}_{Maxwellian}$ is shown in
Fig.\ref{fig:compatibility_all}(a), where regions in the
$m_{\chi}$--$\delta$ plane enclosed by lines represent configurations
with ${\cal D}_{\mbox{Halo indep.}}<1$, and shaded areas those with
${\cal D}_{Maxwellian}<1$ (Fig. \ref{fig:compatibility_all}(b) is a
zoom--up of Maxwellian regions). Notice that, as far as the
compatibility factor is concerned, only the nuclear scaling law is
relevant, and so the relativistic models of Table \ref{table:ref_nref}
which correspond to the same non--relativistic effective operator in
Eq. (\ref{eq:H}) lead to identical results. Explicitly: ${\cal
  O}_4^{AA}\rightarrow {\cal O}_4$, ${\cal O}_9^{VA}$, ${\cal
  O}_9^{TA}$, ${\cal O}_9^{AT}\rightarrow {\cal O}_9$, ${\cal
  O}_{10}^{SP}\rightarrow {\cal O}_{10}$, ${\cal O}_6^{PP}$, ${\cal
  O}_6^{T'T'}\rightarrow {\cal O}_6$.  For this reason the relevant
discussion is in terms of non--relativistic operators. Indeed, for all
of them Fig.\ref{fig:compatibility_all} shows that regions in the
parameter space exist where the DAMA effect can be explained in terms
of the particle $\chi_1$ spin--dependent upscatters off protons in
sodium in compliance with all the other constraints. In the same
figure the open circle represents a benchmark for which measurements
and bounds for the functions $\tilde{\eta}_0$, $\tilde{\eta}_1$ are
shown in detail in Fig.\ref{fig:c4_eta_vmin} when the DAMA effect is
interpreted in terms of the non--relativistic operator ${\cal O}_4$.

Actually, as far as the standard spin--dependent coupling ${\cal O}_4$
is concerned, the region of Fig. \ref{fig:compatibility_all}
represents an update of the analysis of Ref.\cite{noi_idm_spin}, since
we include the latest experimental constraints from
PICO-2L\cite{pico2l,pico2l_2016},
LUX\cite{lux,lux_2015_reanalysis,lux_complete} and
PANDA\cite{panda_run_8,panda_run_9}. Notice that new constraints do
not affect the viability of our scenario, thanks to its intrinsic
xenon--phobic and fluoro-phobic nature. On the other hand, the regions
corresponding to the non--standard interactions ${\cal O}_9$, ${\cal
  O}_{10}$ and ${\cal O}_6$ represent an extension of the analysis of
Ref. \cite{noi_idm_spin}. A first feature that is worth pointing out
is that the models ${\cal O}_9$ and ${\cal O}_{10}$ have similar
behaviors.  This is due do the fact that, as can be seen from Table
\ref{table:ref_nref}, they both scale with the second power of the
exchanged momentum $q$, and they depend on the two nuclear response
functions $W^{\tau\tau^{\prime}}_{\Sigma^{\prime}}(q^2)$ and
$W^{\tau\tau^{\prime}}_{\Sigma^{\prime\prime}}(q^2)$, which are
related to the two projections of the nuclear spin along the direction
perpendicular or parallel to the exchanged momentum, respectively,
with the property $W^{\tau\tau^{\prime}}_{\Sigma^{\prime}}(q^2) \simeq
2 W^{\tau\tau^{\prime}}_{\Sigma^{\prime\prime}}(q^2)$ when
$q\rightarrow$ 0, implying a similar scaling law of the cross section
off different nuclei. Moreover, the allowed regions shrink when
passing from ${\cal O}_4$ to ${\cal O}_9\simeq {\cal O}_{10}$ and
finally to ${\cal O}_6$, both for a halo--independent and a Maxwellian
analysis. Such behavior is easily understood when one considers that
our DM candidate does not interact with xenon, germanium and fluorine,
and that for its relevant masses, $m_{\chi}\lsim$ 20 GeV, only iodine
in COUPP (with a recoil energy threshold $E_{th}$=7.8 keV which is
smaller than that in DAMA and KIMS) is sensitive enough to put
constraints on it. We have confirmed this by numerical
inspection. Then, scattering off sodium implies smaller exchanged
momenta compared to those for scattering off iodine (notice that
$q=\sqrt{2 M_N E_R}$), so a larger power of $q$ implies a relative
suppression of the expected DAMA signal compared to the same quantity
in COUPP, and so a more stringent limit and a smaller allowed
parameter space. Indeed for ${\cal O}_4$, ${\cal O}_9\simeq {\cal
  O}_{10}$ and ${\cal O}_6$ the nuclear response function scales as
$q^{n}$, with $n=0,2,4$, respectively.

\begin{figure}
\begin{center}
\includegraphics[width=0.49\columnwidth, bb=64 197 505 630]{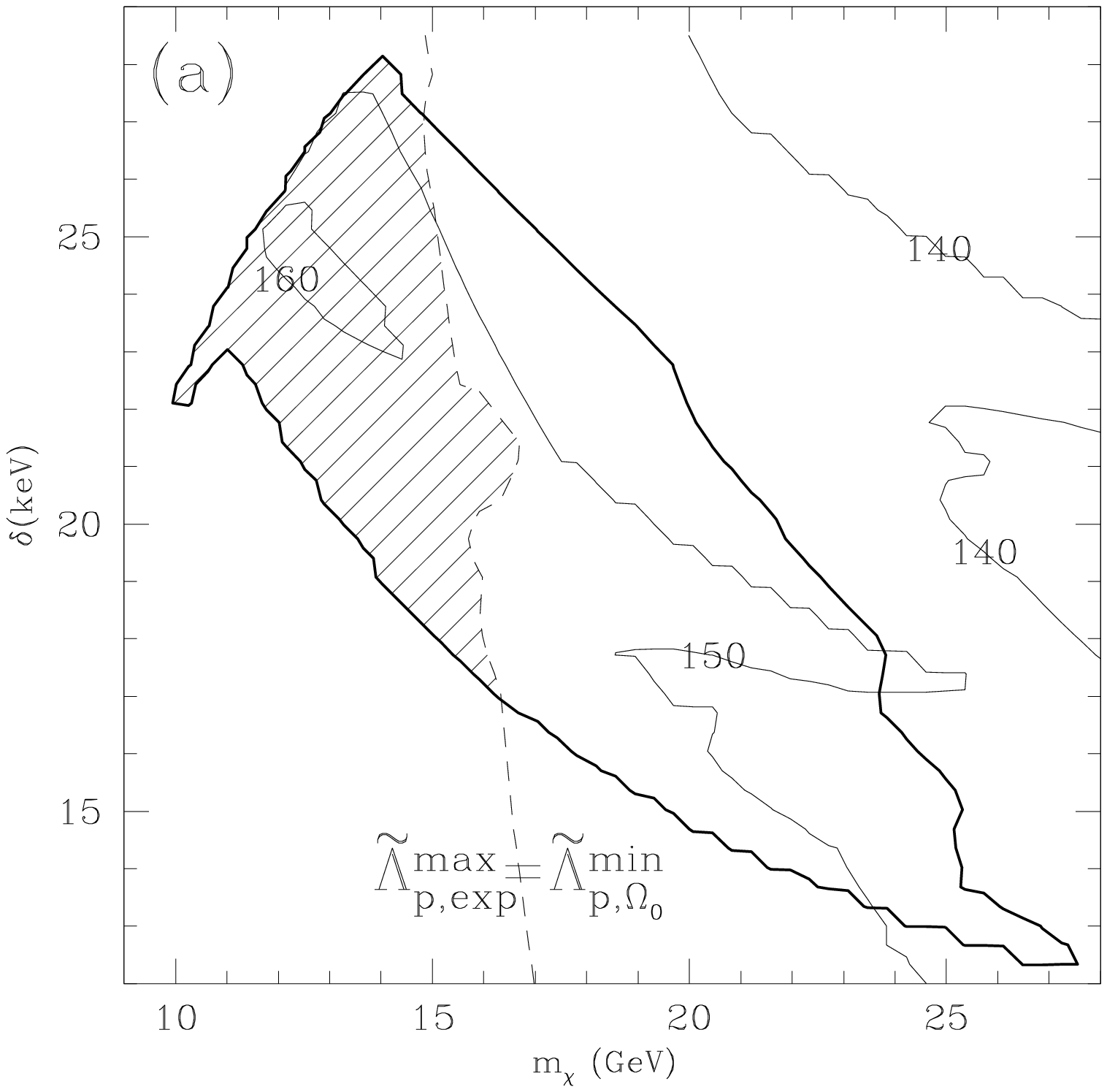} 
\includegraphics[width=0.49\columnwidth, bb=64 197 505 630]{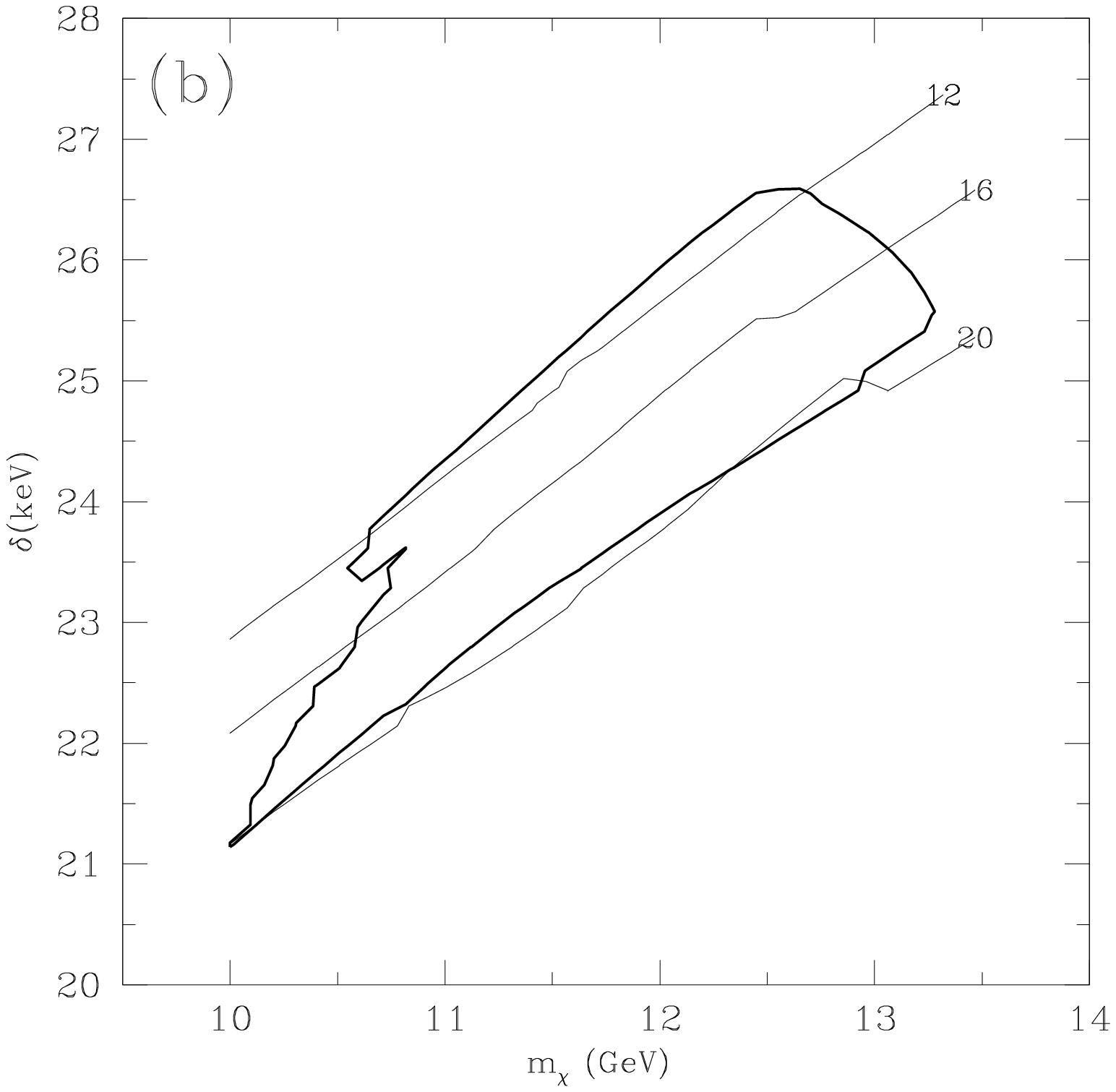} 
\end{center}
\caption{In each plot the thick solid contour encloses the region in
  the $m_{\chi}$--$\delta$ parameter space where an explanation of the
  DAMA modulation effect in terms of the effective interaction ${\cal
    O}^{AA}_4$ yields a compatibility factor ${\cal D}<$1; thin solid
  lines show constant values of the effective scale
  $\tilde{\Lambda}^{max}_{p,exp}$, as indicated by the labels. {\bf
    (a)} Halo--independent case, ${\cal D}$=${\cal D}_{\mbox{Halo
      indep.}}$ (see
  Eq. (\ref{eq:compatibility_factor_generilized})). The dashed line
  indicates the configurations where
  $\tilde{\Lambda}^{max}_{p,exp}$=$\tilde{\Lambda}^{min}_{p,\Omega_0}$,
  and the dashed region where
  $\tilde{\Lambda}^{max}_{p,exp}>\tilde{\Lambda}^{min}_{p,\Omega_0}$;{\bf
    (b)} WIMP galactic velocity distribution fixed to a Maxwellian,
  ${\cal D}$=${\cal D}_{Maxwellian}$ (see
  Eq.(\ref{eq:compatibility_factor_maxwellian})).}
\label{fig:c4_lambda_p_tilde_exp}
\end{figure}

\begin{figure}
\begin{center}
\includegraphics[width=0.49\columnwidth, bb=64 197 505 630]{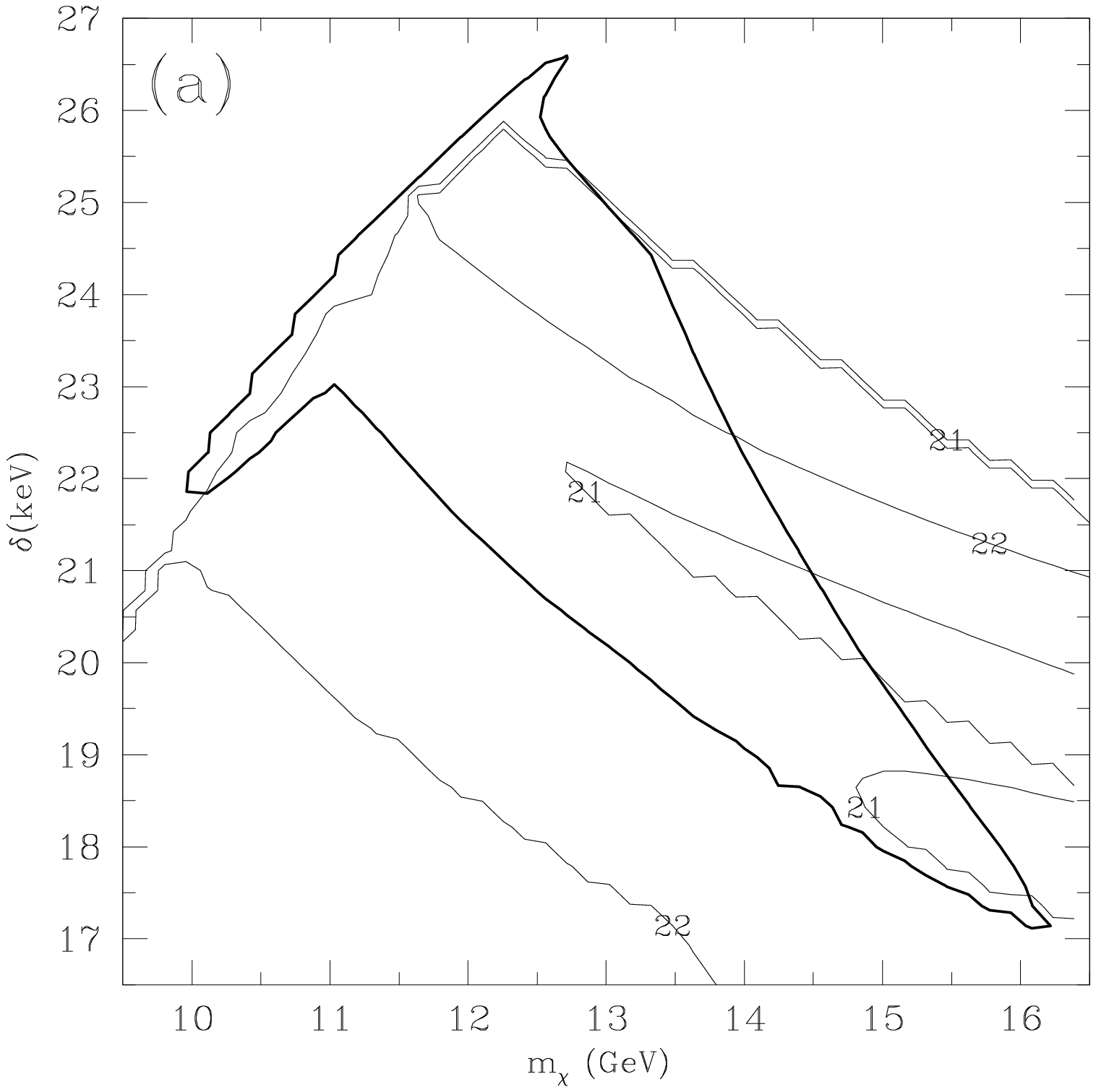} 
\includegraphics[width=0.49\columnwidth, bb=64 197 505 630]{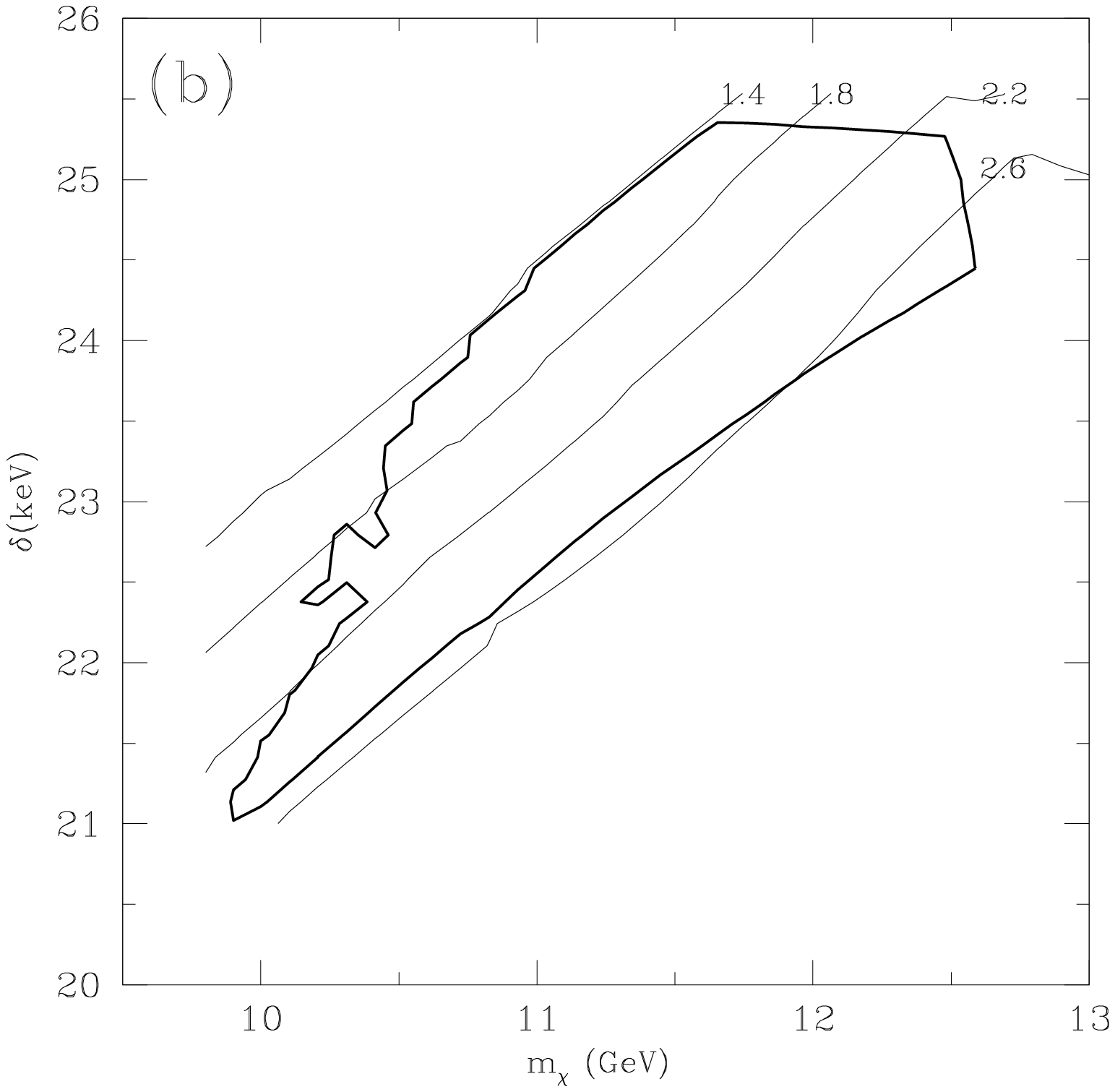} 
\end{center}
\caption{The same as in Fig.\ref{fig:c4_lambda_p_tilde_exp} for the
  non--relativistic operator ${\cal O}_9$, corresponding to the
  relativistic models ${\cal O}_9^{VA}$, ${\cal O}_9^{TA}$ and ${\cal
    O}_9^{AT}$.}
\label{fig:c9_lambda_p_tilde_exp}
\end{figure}

As explained in Section \ref{sec:bottom_up}, a discussion of the
thermal relic abundance requires the determination from experimental
data of the effective scale $\tilde{\Lambda}^{max}_{p,exp}$ defined in
Eq. (\ref{eq:lambda_tilde_p}). As a way of example, contour plots of
$\tilde{\Lambda}^{max}_{p,exp}$ are shown in
Figs. \ref{fig:c4_lambda_p_tilde_exp} and
\ref{fig:c9_lambda_p_tilde_exp} for two of the non--relativistic
operators, ${\cal O}_4$ and ${\cal O}_9$, respectively. In both
figures the left--handed plot shows the result of the
halo--independent procedure described in
Eq. (\ref{eq:lambda_tilde_p_halo_indep}), while the right--handed one
corresponds to the same quantity when the WIMP velocity distribution
is fixed to a Maxwellian. The situation for all the non--relativistic
couplings is summarized in Table \ref{table:lambda_p_tilde_max}, where
the maximal value reached by $\tilde{\Lambda}^{max}_{p,exp}$ within
the corresponding experimentally viable parameter region ($\cal D<$1)
is shown for a halo--independent and a Maxwellian analysis. A first
observation is that in the halo--independent case the effective scale
turns out to be larger than in the Maxwellian one, an obvious
consequence of the choice for the modulated halo function
$\tilde{\eta}_1$ of the expression given in Eq.(\ref{eq:eta_max_vs}):
such functional form is numerically much larger than the one in the
case of a Maxwellian, and leads to a smaller $\sigma_{ref}$ and a
larger $\tilde{\Lambda}^{max}_{p,exp}$. The second feature worth to be
pointed out is that, since non--standard interactions with explicit
momentum dependence have response functions proportional to powers of
factors $(q/m_{\cal N})^2<<$1 (with $m_{\cal N}$ the nucleon mass, see
Table. (\ref{table:ref_nref})), the ensuing suppression in the
expected rate off sodium needs to be compensated by smaller
$\tilde{\Lambda}^{max}_{p,exp}$ values to explain the observed
modulation amplitudes in DAMA. This explains why in Table
\ref{table:lambda_p_tilde_max} the effective scale
$\tilde{\Lambda}^{max}_{p,exp}$ decreases moving from ${\cal O}_4$ to
${\cal O}_9\simeq {\cal O}_{10}$ and finally to ${\cal O}_6$.  Notice
that strictly speaking this is a spurious effect since in some of the
interactions terms of Table \ref{table:ref_nref} the choice
$m_M$=$m_{\cal N}$ in the factors $(q/m_M)$ is arbitrary. However the
choice of the scale $m_M$ affects direct detection and the relic
abundance in the same way, and so does not modify the correlation
among the two quantities that is the object of our study.

Moving on to such correlation, the conclusion that can be drawn from
Figs.\ref{fig:c4_lambda_p_tilde_exp}, \ref{fig:c9_lambda_p_tilde_exp}
and Table \ref{table:lambda_p_tilde_max} is that, when the
corresponding ranges of $\tilde{\Lambda}^{max}_{p,exp}$ are compared
to the curves for $\tilde{\Lambda}^{min}_{p,\Omega_0}$ shown in
Fig.\ref{fig:lambda_p_tilde_min} the inequality
$\tilde{\Lambda}^{max}_{p,exp} > \tilde{\Lambda}^{min}_{p,\Omega_0}$
(see Eq.(\ref{eq:relic_condition})) can be only verified for the
halo--independent analysis of the ${\cal O}_4$ model. Indeed, in all
other cases $\tilde{\Lambda}^{min}_{p,\Omega_0}$ needs to be too large
to get $\Omega$=$\Omega_0$, an indication that the combination $\sum_q
N_q c_q^2 \tilde{<\sigma>}_q$ in the annihilation cross section (see
Eq.(\ref{eq:sigmav1})) is also too large. Notice that this implies
that our choice to include in the relic abundance calculation only the
couplings of the DM particles to the quarks that drive spin--dependent
direct detection already includes the most favorable situation and
adding more couplings would not change our conclusions. The same
argument justifies our choice $|c_{\nu}|\ll |c_q|$ at the end of
Section \ref{sec:relic}. 

Notice that in Table \ref{table:lambda_p_tilde_max} for the models
that do not verify the condition of Eq. (\ref{eq:relic_condition}) the
validity of EFT at the scale of thermal decoupling, $\Lambda\gsim 2
m_{\chi}$, needs not to be a requirement, since anyway some
alternative mechanism, involving additional degrees of freedom, needs
to be advocated for the relic abundance. Such models can explain the
DAMA effect provided that the relic abundance is fixed to the observed
one by some other mechanism, and that the EFT is valid at the much
lower scale of the direct detection process, q$\lsim$100 MeV/c.  We
find that, when the corresponding values of the effective scale
$\tilde{\Lambda}^{max}_{p,exp}$ are converted to the physical cut--off
scale $\Lambda$ using Eq. (\ref{eq:lambda_tilde_p}) such requirement
can be met by all the models of Table 2. In particular, for the most
favorable choice of parameters in Eq. (\ref{eq:lambda_tilde_p})
($\xi=1$ and $|c_p|^{1/2}$ maximized in terms of the coupling
constants) $\Lambda$ is typically of the order of a few tens of GeV,
and always larger that about 800 MeV in the most extreme cases.

As far as model ${\cal O}_4$ is concerned, in
Fig.\ref{fig:c4_lambda_p_tilde_exp}(a) the dashed line represents
configuration where
$\tilde{\Lambda}^{max}_{p,exp}$=$\tilde{\Lambda}^{min}_{p,\Omega_0}$,
and the dashed region where
$\tilde{\Lambda}^{max}_{p,exp}>\tilde{\Lambda}^{min}_{p,\Omega_0}$. Such
region represents the only pSIDM parameter space interval, including
the generalizations of Table \ref{table:ref_nref}, compatible to the
assumption that the same effective dimension--six operator that
explains the DAMA excess can also yield an acceptable relic abundance
through thermal decoupling. To be more specific, in such parameter
space $\tilde{\Lambda}^{min}_{p,\Omega_0}$ is achieved for $r_d\simeq$
0.47 and $r_s\simeq$ -0.3 which, at fixed values of $m_{\chi}$ can be
converted in explicit allowed intervals of the couplings $c_q$ and the
cut--off scale $\Lambda$ using Eq.(4.16) and the relevant curve of
Fig. 1. For instance, for $m_{\chi}\simeq$ 15 GeV one has
$\xi_{min}\simeq$ 1.5$\times$10$^{-6}$, implying:

\begin{equation}
  \left (\frac{\Lambda}{30 \mbox{GeV}} \right )^2 \frac{4\pi}{|c_u|}\lsim 800.
\end{equation}

\noindent A combination fulfilling the relation above is, for instance,
$c_u\simeq$ 1.3, $c_d\simeq$ 0.66, $c_s\simeq$ -0.42 and
$\Lambda\simeq$ 270 GeV.

The allowed parameter space of model ${\cal O}_4^{AA}$ would shrink if
additional couplings besides the minimal ones driving direct detection
were included in the relic abundance calculation, and for a halo
function $\tilde{\eta}_1$ different from Eq.(\ref{eq:eta_max_vs}). On
the other hand no compatibility is achieved in the Maxwellian case
irrespective of the EFT model.

\begin{table}[t]
\begin{center}
{\begin{tabular}{@{}|c|c|c|@{}}
\hline
Non--relativistic operator  &  Halo--independent  & Maxwellian\\
\hline\hline
${\cal O}_4$ & 163   & 20.4\\
${\cal O}_6$ & 23.0   & 2.79\\
${\cal O}_9$ & 27.5   & 3.27\\
${\cal O}_{10}$ & 3.01   & 0.32\\
\hline
\end{tabular}}
\caption{Maximal values in GeV of the effective scale
  $\tilde{\Lambda}^{max}_{p,exp}$ within the parameter space region
  where the compatibility factors ${\cal D}_{\mbox{Halo indep.}}<1$ or
  ${\cal D}_{Maxwellian}<1$ for each of the non--relativistic
operators of Eq.(\ref{eq:H}).
  \label{table:lambda_p_tilde_max}}
\end{center}
\end{table}

\begin{figure}
\begin{center}
\includegraphics[width=0.49\columnwidth, bb=64 197 505 630]{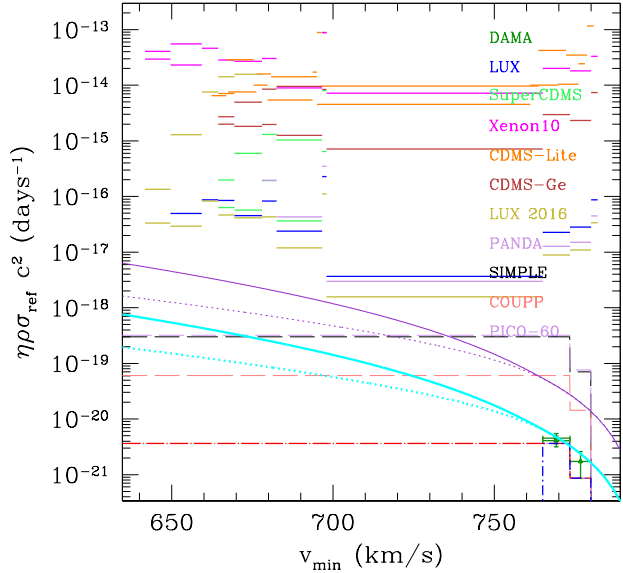} 
\end{center}
\caption{Measurements and bounds for the functions $\tilde{\eta}_0$,
  $\tilde{\eta}_1$ defined in Eqs.(\protect\ref{eq:eta_tilde}) for the
  non--relativistic coupling ${\cal O}_4$ and $m_{\chi}$=12 GeV,
  $\delta$=25 keV (the benchmark point $P$ indicated with an open
  circle in Fig.\ref{fig:compatibility_all}). The (green) triangles
  represent the $\tilde{\eta}_1$ estimations from DAMA, while
  horizontal lines show upper bounds from other experiments, as
  indicated in the plot. The thick (cyan) dotted line represents the
  best--fit on DAMA points of the function $\tilde{\eta}_1$ in the
  case of a Maxwellian distribution with $v_{esc}$=550 km/sec and
  $v_{rms}$=270 km/sec, while the thick (cyan) solid line is the
  corresponding $\tilde{\eta}_0$. On the other hand the (purple) thin
  solid line represents the maximal $\tilde{\eta}$ for the Maxwellian
  case allowed by COUPP, while the (purple) thin dotted line is the
  corresponding $\tilde{\eta}_1$. The (blue) dots -- short dashes
  represent a conservative piece-wise estimation of the function
  $\tilde{\eta}^{est}_1(v_{min})$ passing through DAMA points: (red)
  dots -- long dashes show the corresponding minimal piece-wise
  estimation of the function $\tilde{\eta}_0^{est}(v_{min})$ (in
  compliance to the requirements of Eq.(\ref{eq:eta_conditions})) used
  to calculate the quantities $N_{k,i}^{expected}$ of
  Eq.(\ref{eq:compatibility_factor_generilized}) for droplet detectors
  and bubble chambers. The lower long--dashed (black) line shows the
  maximal $\tilde{\eta}^{est}(v_{min})$ allowed by scatterings off
  chlorine in SIMPLE, while the upper (purple) one is the same curve
  for scatterings off fluorine in PICO-60.}
\label{fig:c4_eta_vmin}
\end{figure}

\section{Conclusions}
\label{sec:conclusions}

In the present paper we have discussed strategies to make inferences
on the thermal relic abundance of a Weakly Interacting Massive
Particle (WIMP) when the same effective dimension--six operator that
explains an experimental excess in direct detection is assumed to
drive decoupling at freeze--out, and applied them to the explicit
pSIDM scenario introduced in Ref. \cite{noi_idm_spin}, a
phenomenological set--up containing two DM states $\chi_1$ and
$\chi_2$ with spin--dependent couplings to protons, masses
$m_{\chi}$=$m_{\chi_1}<m_{\chi_2}$ and a small mass splitting $\delta$
that we showed to explain the DAMA effect in compliance with the
constraints from other detectors. Compared to Ref. \cite{noi_idm_spin}
we have updated experimental constraints including the latest bounds
from LUX\cite{lux_2015_reanalysis,lux_complete},
PANDA\cite{panda_run_8,panda_run_9} and PICO-2L\cite{pico2l_2016}, and
extended the analysis to the most general spin--dependent
momentum--dependent interactions allowed by non--relativistic
effective theory, listed in Table \ref{table:ref_nref}.

To assess the viability of an interpretation of the DAMA modulation
effect in terms of pSIDM we have calculated the compatibility factors
${\cal D}_{\mbox{Halo indep.}}$ of
Eq.(\ref{eq:compatibility_factor_generilized}) and ${\cal
  D}_{Maxwellian}$ of Eq.(\ref{eq:compatibility_factor_maxwellian}):
the resulting allowed regions, defined by the requirements ${\cal
  D}_{\mbox{Halo indep.}}<1$ and ${\cal D}_{Maxwellian}<1$, are shown
in Fig.\ref{fig:compatibility_all}. The largest allowed parameter
space is obtained for the standard spin--dependent ${\cal O}_4$
non--relativistic coupling and for a halo--independent analysis, 10
GeV $\lsim m_{\chi}\lsim $ 27.5 GeV, 12.3 keV $\lsim \delta \lsim $
28.2 keV, although for all the models of Table \ref{table:ref_nref}
compatible regions exist, also when a Maxwellian velocity distribution
is assumed for the WIMPs, albeit in smaller parameter ranges.

The bottom--up problem of calculating the relic abundance (at the GeV
scale) fixing the model parameters with direct detection data (at the
keV scale) represents the first necessary step upward in the
bottom--up quest to shed light on the ultraviolet completion of the
EFT. As discussed in Section \ref{sec:analysis}, it is affected by a
degeneracy between the WIMP local density $\rho_{\chi}$ and the
effective WIMP--nucleon scattering cross section $\sigma_{ref}$ that
parameterizes the particle--physics interaction, since $\rho_{\chi}$
must be rescaled with respect to the observed DM density in the
neighborhood of the Sun when the calculated relic density $\Omega$ is
smaller than the observed one $\Omega_0$. As a consequence, a direct
detection experiment is not directly sensitive to the physical
cut--off scale $\Lambda$ of the effective theory, but on the
dimensional combination $\tilde{\Lambda}_{p,exp}$ defined in
Eq. (\ref{eq:lambda_tilde_p}) and predicted by a given effective
theory to be equal to the quantity $\tilde{\Lambda}_{p,th}$ defined in
Eq.(\ref{eq:lambda_tilde_p_th}) and plotted in
Fig. \ref{fig:lambda_p_tilde_min} for the scenarios of Table
\ref{table:ref_nref}.  Although $\tilde{\Lambda}_{p,th}$ is
numerically equal to the interaction strength
$\tilde{\Lambda}_{p,\Omega_0}$ required to get the observed relic
abundance $\Omega_0$, the dependence on the physical scale $\Lambda$
cancels out in its calculation, so it is not correlated to the actual
$\Omega$. In other words, due to rescaling a direct detection
experiment is bound to measure
$\tilde{\Lambda}_p=\tilde{\Lambda}_{p,\Omega_0}$ even if
$\Omega\ll\Omega_0$. The only direct test on the physical scale is the
upper bound on $\Lambda$ corresponding to the condition $\Omega\le
\Omega_0$, which should be compared to the lower bound $ \Lambda\gsim
2 m_{\chi}$ required for the Effective Theory to be valid and to
$|c_q|\le 4\pi$ from perturbativity; for the specific scenarios discussed
in our paper such constraint turns out to be irrelevant (see
Fig.\ref{fig:csi_min}).

The degeneracy discussed above allows to develop the consistency test
of Eq. (\ref{eq:relic_condition}) for a given experimental excess:

\begin{equation}
  \tilde{\Lambda}_{p,exp}^{max}>\tilde{\Lambda}_{p,\Omega_0}^{min},
  \nonumber
\end{equation}

\noindent where $\tilde{\Lambda}_{p,\Omega_0}$ is minimized with
respect to the couplings ratios $r_q=c_q/c_u$, at fixed WIMP--neutron
coupling $c_n$=0 (as required by the proton--philic nature of our DM
candidate). Such test is not simply a requirement to obtain the
correct relic abundance or to avoid overclosure ($\Omega>\Omega_0$)
when the same parameters that explain a given experimental excess are
used in the coannihilation cross section. Instead, it should be
interpreted as a more general condition on the possibility that the
same effective model that explains such excess can also determine
through thermal decoupling the relic abundance $\Omega$ in a
consistent way, and irrespective of the actual value of $\Omega$.

In Section \ref{sec:analysis} we have calculated in a systematic way
the scale $\tilde{\Lambda}_{p,exp}^{max}$ from the DAMA data,
summarizing the result in Figs. \ref{fig:c4_lambda_p_tilde_exp} and
\ref{fig:c9_lambda_p_tilde_exp} and in Table
\ref{table:lambda_p_tilde_max}. In particular, when comparing for each
model the maximum value reached by $\tilde{\Lambda}_{p,exp}^{max}$ as
provided in Table \ref{table:lambda_p_tilde_max} to the curves for
$\tilde{\Lambda}^{min}_{p,\Omega_0}$ shown in
Fig.\ref{fig:lambda_p_tilde_min} it is possible to conclude that the
inequality $\tilde{\Lambda}^{max}_{p,exp} >
\tilde{\Lambda}^{min}_{p,\Omega_0}$ can be only verified for the
halo--independent analysis of a standard spin--dependent interaction
${\cal O}$=$\bar{\chi}_1\gamma^{\mu}\gamma^5\chi_2
\bar{q}\gamma_{\mu}\gamma^5 q$ +h.c., with no explicit momentum
dependence. The ensuing allowed parameter space corresponds to the
shades area of Fig.\ref{fig:c4_lambda_p_tilde_exp}(a), i.e.,
approximately, 10 GeV $\lsim m_{\chi}\lsim$ 16 GeV, 17 keV$\lsim
\delta\lsim$28 keV. From Fig.\ref{fig:csi_min} one can see that the
actual value of $\Omega$ is basically undetermined in this scenario,
$6\times 10^{-7}\Omega_0\lsim \Omega \lsim \Omega_0$. Such degeneracy
on the physical cut--off scale $\Lambda$ could in principle be at
least partially resolved by using constraints from accelerator
signals.  For all the other spin--dependent effective models listed in
Table \ref{table:ref_nref} the pSIDM scenario cannot explain the DAMA
effect and at the same time provide a thermal relic if the same
dimension--six operator drives both direct detection and decoupling at
freeze--out, and this conclusion is valid for any choice of the WIMP
galactic velocity distribution $f(v)$. Moreover, in no case the same
effective coupling can provide a thermal relic and explain DAMA
if $f(v)$ is a Maxwellian.

An additional complication of the pSIDM scenario is that for all the
models of Table \ref{table:ref_nref}, if no other coupling is assumed
for the DM particle, the lifetime of the heavy state $\chi_2$ is
larger than the age of the Universe. In such case our Galaxy would be
formed by a mixture of $\chi_1$ and $\chi_2$.  Presumably such states
would be present in equal parts and have the same velocity
distribution, so that direct detection experiments would be both
sensitive to the upscatters of $\chi_1$'s and the downscatters of
$\chi_2$'s. However, as explained in section
\ref{sec:spin_idm_scenario} downscatters of the $\chi_2$ particle off
fluorine must be suppressed in order to comply with the constraints
from bubble chambers and droplet detectors. In our analysis we have
assumed the existence of some additional mechanism (such as the
coupling of Eq.(\ref{eq:l_neutrino}) with $|c_{\nu}|\ll |c_q|$) that
depletes the $\chi_2$ particles without affecting the relic
density. As a consequence, our model does not produce any indirect
detection signal.

\acknowledgments This research was supported by the Basic Science
Research Program through the National Research Foundation of
Korea(NRF) funded by the Ministry of Education, grant number
2016R1D1A1A09917964.  S.S. acknowledges the hospitality of the
Instituto de Fis\'ica Te\'orica (IFT), Madrid, where part of this work
was carried out, and partial support by the Programme SEV-2012-0249
`Centro de Excelencia Severo Ochoa'.

\appendix

\section{Thermal coannihilation cross sections}
\label{app:sigmav}

For a Cold Dark Matter particle the thermal average of the
coannihilation cross section times velocity at temperature $T$ can be
expressed as $<\sigma v>=a+b/x$ with $x\equiv T/m_{\chi}\ll 1$. The
relic abundance is inversely proportional to the combination
$\tilde{<\sigma v>}=a+b/(2 x_f)$ with $x_f\equiv T_f/m_{\chi}\simeq
1/20$ and $T_f$ the freeze--out temperature. Since $T>>\delta$ in the
calculation the mass difference between $\chi_1$ and $\chi_2$ can be
safely neglected. The coefficients $a$ and $b$ for the different
models discussed in the present paper are given by $a=1/\Lambda^4
\sum_f N_f c_f^2 (a)_f$, $b=1/\Lambda^4 \sum_f N_f c_f^2 (b)_f$, where
the sum is over the final fermionic states $f$ with coupling
$c_f^2/\Lambda^4 \bar{\chi}_1\Gamma^a\chi_2 \bar{f} \Gamma^{b} f$ and
internal degrees of freedom $N_f$, and, using the same notation of
Table \ref{table:ref_nref} to indicate the different effective models
considered in the present paper, $(a)_f$ and $(b)_f$ are given by:

\begin{equation}
(a^{AA}_{4})_f=
\frac{m_f^2 \sqrt{m_{\chi}^2-m_f^2}}{2 \pi  m_{\chi}};
\;\;\;\;
(b^{AA}_{4})_f=
\frac{23 m_f^4-28 m_f^2 m_{\chi}^2+8 m_{\chi}^4}
{8 \pi m_{\chi} \sqrt{m_{\chi}^2-m_f^2}}
\nonumber
\end{equation}

\begin{equation}
(a^{VA}_{9})_f=
  \frac{(m_{\chi}^2-m_f^2)^{3/2}}{\pi m_{\chi}};
\;\;\;\;
(b^{VA}_{9})_f=
  \frac{(11 m_f^2-2 m_{\chi}^2) \sqrt{m_{\chi}^2-m_f^2}}{4 \pi m_{\chi}}
  \nonumber  
\end{equation}

\begin{equation}  
(a^{TA}_{9})_f=
  \frac{4 m_{\chi} (m_{\chi}^2-m_f^2)^{3/2}}{\pi m_{\cal N}^2};
  \;\;\;\;
(b^{TA}_{9})_f=
  \frac{m_{\chi} \sqrt{m_{\chi}^2-m_f^2} (7 m_f^2+2 m_{\chi}^2)}
  {\pi m_{\cal N}^2}  
\nonumber  
\end{equation}

\begin{equation}
(a^{AT}_{9})_f=0;
  \;\;\;\;
(b^{AT}_{9})_f=   
   \frac{2 m_{\chi}
    \sqrt{m_{\chi}^2-m_f^2} (2
    m_f^2+m_{\chi}^2)}{\pi m_{\cal N}^2}
 \nonumber
\end{equation}

\begin{equation}
(a^{SP}_{10})_f=0;
  \;\;\;\;
(b^{SP}_{10})_f=
  \frac{3 m_{\chi} \sqrt{m_{\chi}^2-m_f^2}}{4 \pi}
\nonumber  
\end{equation}

\begin{equation}  
(a^{PP}_{6})_f=
  \frac{m_{\chi} \sqrt{m_{\chi}^2-m_f^2}}{2 \pi };
\;\;\;\;  
(a^{PP}_{6})_f= \frac{3 m_f^2 m_{\chi}}{8 \pi \sqrt{m_{\chi}^2-m_f^2}}
\nonumber  
\end{equation}

\begin{equation}
 (a^{T'T'}_{6})_f=0;
  \;\;\;\;
(b^{T'T'}_{6})_f= 
  \frac{4 (m_{\chi}^4-m_f^2 m_{\chi}^2)^{3/2}}{\pi m_{\cal N}^4}
\nonumber  
\end{equation}
\noindent The observed DM density corresponds to the reference
value $\tilde{<\sigma v>}_0=2\times 10^{-9}$ GeV$^{-2}$ (in natural
units), and the rescaling factor discussed in section
\ref{sec:bottom_up} is given by $\xi=\tilde{<\sigma
  v>}_0/\tilde{<\sigma v>}$.

\section{Nuclear matrix elements}
\label{app:matrix_elements}

As, for instance, summarized in Section A of \cite{dienes} it is
usually assumed that between the WIMP--nucleon and the WIMP--quark
bilinear currents the following relation holds:

\be
<{\cal N}_f|\bar{q}\Gamma^I q|{\cal N}_i>\equiv \Delta q^{{\cal N}} <{\cal N}_f|\bar{{\cal N}}\Gamma^I {\cal N}|{\cal N}_i>,
\label{eq:matrix_elements}
\ee

\noindent with $\Delta q^{{\cal N}}$ quantities incorporating the
complicated non--perturbative physics through which quarks are
confined into the nucleons ${\cal N}=n,p$. Usually, such quantities,
which most of the times cannot be calculated from first principles but
must be estimated from some measured processes depending on the same
current, are taken as constants, i.e. independent on the spins and
velocities of the nucleon and the quark.

Once the $\Delta q^{{\cal N}}$'s are known, it is straightforward to
calculate the effective WIMP--nucleon couplings in terms of the
WIMP--quark couplings:

\begin{equation}
c_{\cal N}=K(\sum_q c_q \Delta q^{{\cal N}}).
\label{eq:couplings_qp}
\end{equation}
\noindent where the constant $K$ can be directly read--off comparing
the second and fourth columns in Table \ref{table:ref_nref}.

\noindent In particular the axial current $\Gamma^A$ is directly
related to the nucleon spin (actually, the corresponding $\Delta
q^{{\cal N}}_A$'s can be directly interpreted as the fractions of the
nucleon spin carried by the quark $q$). On the other hand the
pseudoscalar current $\Gamma_5$ can be related to $\Gamma^A$ due to
the fact that the latter is not conserved in the nucleon for
non--vanishing quark masses \cite{cheng_bai1,cheng_bai2,cheng_bai3} so
that the corresponding $\Delta q^{{\cal N}}_5$ are related to the
$\Delta q^{{\cal N}}_A$'s\cite{dienes}. The numerical values of both
the $\Delta q^{{\cal N}}_A$ and $\Delta q^{{\cal N}}_5$'s are for
instance given in \cite{krauss} and, for $q$=$u$, $d$, $s$ are shown
in Tables \ref{table:axial} and \ref{table:pseudoscalar}.

\begin{table}[t]
\begin{center}
{\begin{tabular}{@{}|c|c|c|@{}}
\hline
{\bf $\Delta q^{{\cal N}}_A$} &  $n$  & $p$ \\
\hline
  u      & -0.427  & 0.842 \\
  d      & 0.842  & -0.427 \\
  s      & -0.085  & -0.085 \\ 
\hline
\end{tabular}}
\caption{$\Delta q^{{\cal N}}_A$ estimations of the axial coupling
  $\Gamma^A$=$\gamma^{\mu}\gamma_5$ for $q=u,d,s$.
  \label{table:axial}}
\end{center}
\end{table}

\begin{table}[t]
\begin{center}
{\begin{tabular}{@{}|c|c|c|@{}}
\hline
{\bf $\Delta q^{{\cal N}}_5$} &  $n$  & $p$ \\
\hline
  u      & -108.3  & 110.55 \\
  d      & 108.60  & -107.17 \\
  s      & -0.57  & -3.37 \\ 
\hline
\end{tabular}}
\caption{The same of Table \ref{table:axial} for $\Delta q^{{\cal N}}_5$.
  \label{table:pseudoscalar}}
\end{center}
\end{table}

\begin{table}[t]
\begin{center}
{\begin{tabular}{@{}|c|c|c|c|c|@{}}
\hline
{\bf $\Delta q^{p}_T$=$\Delta q^{p}_{T'}$} & HERMES\cite{transversity_hermes}  & COMPASS\cite{transversity_compass} & Dyson--Schwinger\cite{transversity_sr,yamanaka} & Lattice\cite{transversity_lattice} \\
\hline
u  & 0.57(21)  & 0.39$^{+0.18}_{-0.12}$  & 0.55(8) & 0.79(7)  \\
d  & -0.18(33) & -0.25$^{+0.30}_{-0.10}$ & -0.11(2)& -0.22(3) \\
Q$^2$(GeV) & 1.0 & 0.8 & 2 & 2\\
\hline
\end{tabular}}
\caption{Different estimations of the $\Delta q^{p}_T$=$\Delta
  q^{p}_{T'}$ constants for the proton.The last line indicates the
  energy scale of the measurement/calculation.  The corresponding
  quantities for the neutron are obtainable by assuming isospin
  symmetry.
  \label{table:tensor}}
\end{center}
\end{table}

As far as the tensor current $\Gamma^T$ is concerned, it is related to
the transversity distribution of the nucleon \cite{transversity},
which has to do with the contribution of quarks to the projection of
the spin perpendicular to the nucleon's momentum. Only recently the
$\Delta q^{{\cal N}}_T$'s have been measured by the two experiments
Hermes \cite{transversity_hermes} and COMPASS
\cite{transversity_compass}, while for a long time their only
available estimates had been obtained from from Dyson--Schwinger
formalism \cite{transversity_sr,yamanaka} and lattice QCD
\cite{transversity_lattice}.  In all cases they are affected by large
uncertainties, as shown by the numerical values of the $\Delta
q^{p}_T$'s for the proton shown in Table \ref{table:tensor} (the
corresponding ones for the neutron are obtainable by isospin
symmetry). Moreover, thanks to the identity
$\Gamma^{T'}=\sigma^{\mu\nu}\gamma_5$=$(i/2)\epsilon^{\mu\nu\alpha\beta}\sigma_{\alpha\beta}$
one has $\Delta q^{{\cal N}}_{T'}$=$\Delta q^{{\cal N}}_{T}$.  In our
quantitative analysis we used the Hermes determination of the $\Delta
q_T^{\cal N}$ coefficients.

\section{Experimental constraints}
\label{app:exp}

In the present analysis we include the constraints from
LUX\cite{lux,lux_2015_reanalysis,lux_complete},
PANDA\cite{panda_run_8,panda_run_9}, XENON10\cite{xenon10},
CDMS-$Ge$\cite{cdms_ge}, CDMSlite \cite{cdms_lite},
SuperCDMS\cite{super_cdms}, SIMPLE\cite{simple}, COUPP\cite{coupp},
PICASSO\cite{picasso,picasso_final}, PICO-2L\cite{pico2l,pico2l_2016}
and PICO-60\cite{pico60}).  Compared to Refs.\cite{noi_eft_spin} and
\cite{noi_idm_spin} (to which we refer for the details of the other
experiments) in the present paper we include the latest results from
LUX\cite{lux_2015_reanalysis,lux_complete},
PANDA\cite{panda_run_8,panda_run_9} and PICO-2L\cite{pico2l_2016}.

As far as the LUX and PANDA analyses are concerned, we use the
generalization of the response function ${\cal R}$ as given in
Appendix A of Ref.\cite{noi1}.

For LUX we use the events in the lower half of the signal region of
Fig. 1 of Ref. \cite{lux_complete} with a total exposition of
3.35$\times$10$^5$ kg day and the efficiency from Fig.2 of
Ref. \cite{lux_complete} multiplied by 0.5, and convert recoil
energies to photo--electron (PE) numbers using Eq. (3) of
\cite{lux_calibration} with $g_1$=0.115 and the $L_{y}$ function taken
from Fig. 10 of the same paper. 

For PANDA we combine the events in the lower half of the signal
regions of Fig. 4 of \cite{panda_run_8} and Fig. 5 of
\cite{panda_run_9}, assuming a total exposition of 3.3$\times$10$^5$
kg day and the efficiency from Fig.2 of Ref. \cite{panda_run_9}
multiplied by 0.5. The PANDA collaboration does not provide details on
how to convert the recoil energy $E_R$ (in keVnr) into the signal
$S_1$ (in PE). We decide to adopt the same recipe of LUX after
checking that it quantitatively reproduces fairly well the
correspondence between $S_1$ and $E_R$ ranges as deduced by the plots
in \cite{panda_run_9}.

The latest PICO-2L analysis \cite{pico2l_2016} was performed after an
upgrade of the detector that significantly reduced the background
compared to \cite{pico2l}. Only the threshold $E_{th}$=3.3 keV was
analyzed, with a total exposure of 129.0 kg day and 1 event detected
(corresponding to a 95\% upper bound of 5.14 events). We use the
nucleation probability give in Eq.(B.3) of \cite{noi_eft_spin} with
$\alpha_F$=$\alpha_C$=5, and the efficiency of Fig.4 of \cite{pico2l}.


\end{document}